\documentclass{aa}
\usepackage{epsfig}
\usepackage{graphics}
\usepackage{subfigure}
\usepackage{natbib}
\usepackage{amsmath}
\usepackage{amssymb}
\usepackage{enumerate}
\usepackage{threeparttable}

\bibpunct{(}{)}{;}{a}{}{,} % to follow the A&A style

\begin{document}

\title{Collective non-thermal emission from an extragalactic jet interacting with stars}

\author{Florencia L. Vieyro\inst{1}, N\'uria Torres-Alb\`a\inst{1} and Valent\'i Bosch-Ramon\inst{1}}
  
\institute{Departament de F\'{i}sica Qu\`antica i Astrof\'{i}sica, Institut de Ci\`encies del Cosmos (ICC), Universitat de Barcelona (IEEC-UB), Mart\'{i} i Franqu\`es 1, E08028 Barcelona, Spain}

\offprints{F. L. Vieyro \\ \email{fvieyro@fqa.ub.edu}}

\titlerunning{Non-thermal emission from AGN jets}

%\authorrunning{ }

\abstract
{The central regions of galaxies are complex environments, rich in evolved and/or massive stars. For galaxies hosting an active galactic nucleus (AGN) with jets, the interaction of the jets with the winds of the stars within can lead to particle acceleration, and to extended high-energy emitting regions.}
{We compute the non-thermal emission produced by the jet flow shocked by stellar winds on the jet scale, far from the jet-star direct interaction region.}
{First, prescriptions for the winds of the relevant stellar populations in different types of galaxies are obtained. The scenarios adopted include galaxies with their central regions dominated by old or young stellar populations, and with jets of different power. Then, we estimate the available energy to accelerate particles in the jet shock, and compute the transport and energy evolution of the accelerated electrons, plus their synchrotron and inverse Compton emission, in the shocked flow along the jet.}
{A significant fraction of the jet energy, $\sim 0.1-10$\%, can potentially be available for the particles accelerated in jet-wind shocks in the studied cases. The non-thermal particles can produce most of the high-energy radiation on jet scales, far from the jet shock region. This high-energy emission will be strongly enhanced in jets aligned with the line of sight due to Doppler boosting effects.}
{The interaction of relativistic jets with stellar winds may contribute significantly to the persistent high-energy emission in some AGNs with jets. However, in the particular case of M87, this component seems too low to explain the observed gamma-ray fluxes.}
 
 
\maketitle

\section{Introduction}

Active galactic nuclei (AGN) are composed of a supermassive black hole that accretes material from the inner region of the galaxy host. Some AGN are associated with the production of collimated relativistic outflows or jets \citep[e.g.,][]{begelman84}. These jets propagate through complex environments, rich in stars, dust gas, clouds, and even stellar clusters. It is very likely, then, that extragalactic jets interact with the obstacles present in the central region of galaxies. These interactions can affect the jet dynamically on different scales \citep[e.g.,][]{blandford79,wang00,sutherland07,jeyakumar09}. For instance, the penetration of stars with strong winds inside the jet has been proposed as a possible mechanism for jet mass-loading and deceleration \citep[e.g.,][]{komissarov1994,bowman96,hubbard06,perucho2014}. 

In addition to jet dynamical effects, the presence of stars inside the jet can also lead to the generation of high-energy emission. The interaction of a relativistic jet with a powerful stellar wind produces a double bow-shock structure. The shock in the jet flow is a potential site of particle acceleration, and can contribute to the jet non-thermal emission. There have been several works exploring the gamma-ray emission, in the form of both steady radiation and transient events, due to jet-obstacle interactions \citep[e.g.,][]{dar97,bednarek97,beall99,araudo2010}, and in particular, due to jet-star interactions \citep{barkov2010,araudo2013,bednarek2015,moreno2016,banasinski16}. There is also some direct and indirect evidence of jet-star interactions and jet mass-load by stellar winds \citep[e.g.,][and references therein]{muller14,wykes13,wykes15}.

Recent numerical simulations have shown that: (i) the effective surface of the shock induced by an obstacle is larger than the obstacle section, increasing the conversion of kinetic energy into internal energy; and (ii) Doppler boosting has to be taken into account even for standing shocks \citep{bosch-ramon2015,moreno2016}. In addition, it has been found that for jet-star interactions taking place at relatively large distances from the central source, say $\gtrsim$ pc-scale, accelerated particles are not strongly cooled close to the shock. In fact, the non-thermal particles can cover distances similar to the interaction jet height without significant energy loss \citep{bednarek2015,moreno2016}. Regarding the most likely radiation mechanisms, in the jet's innermost regions, hadronic processes cannot be discarded, but in general leptonic emission, namely synchrotron and inverse Compton (IC), will be more efficient in less extreme environments \citep[see][for related discussions]{barkov12a,barkov12b,khangulyan13}.

In this work, we study the collective, steady, leptonic, high-energy radiation resulting from the interaction of an AGN jet with different stellar populations. We compute the non-thermal radiation produced at the scale of the jet, and do not consider in detail the radiation component originated on the smaller scales of the jet-wind interaction structure. The jet-wind interaction region was investigated for individual interactions in \cite{moreno2016},  its emission being roughly generalized for many encounters for the radio galaxy M87 in \cite{bosch-ramon2015}. A detailed study of the extended jet emission as the result of the stellar population in the radiogalaxy Centarus A was conducted by \citet{wykes15}. This study mostly focused on the presence of red giant stars in the host galaxy.

Here, we aim at analyzing the relevance of the jet-scale high-energy emission contribution for different types of galaxy hosts, namely characterized either by old or young stellar components. Our treatment of the problem includes relativistic beaming and accounts for the effective increase in the shock area, which are effects that were not taken into account before when computing the contribution to high-energy emission from collective jet-star interactions on jet scales. We disregard, at this stage, the effects of strong anisotropy in the stellar spatial distribution at the galaxy center, which may influence the number of available stars interacting with the jet.

The article is organized as follows: In Sect. \ref{stellar}, the stellar populations in the central regions of two types of galaxy are characterized; Sect. \ref{model} contains a description of the jet model; in Sect. \ref{hosts}, we outline the properties of three different galaxy hosts ; Sect. \ref{appLnt} presents an estimate of the apparent non-thermal emission for the galaxies studied; whereas Sect. \ref{SED} presents accurate calculations of the transport of relativistic electrons along the jet, and a computation of their high-energy emission. The results are presented in Sect. \ref{results}, and the conclusions in Sect. \ref{discussion}.

\section{Characterization of the stellar populations interacting with AGN jets}\label{stellar}

This work is one of the first approximations to the problem of large-scale emission from jet-star interactions (see also \citealt{bednarek2015}, \citealt{wykes15}), and for this reason two different scenarios are adopted for the types of AGN galaxy host studied: a star-forming galaxy with a dense disk of molecular gas surrounding the nucleus in which the star formation rate (SFR; $\dot{M}_{\rm SFR}$) is very high; and a massive galaxy with an old stellar population distributed in a bulge. More detailed studies of specific sources, or a mixed galaxy with a large population of evolved stars plus a high SFR, are left for future work.

\subsection{Effect on the non-thermal energy budget}

We characterize the stellar populations inside the jet to obtain the luminosity injected in the form of accelerated particles at the jet-star interactions. This non-thermal luminosity can be estimated as
\begin{equation}\label{eq:NTLum}
L_{\rm NT}=\int \int \eta_{\rm NT} \ L_{\rm j} \ \langle \dfrac{S_{\rm s} (m,t)}{S_{\rm j}} \rangle \ n_{\rm s} (m,z) \ {\rm d}m \ {\rm d}z\,,
\end{equation}
where $\eta_{\rm NT}$ is the fraction of jet energy that crosses the effective interaction area $S_{\rm s}$ that is converted into non-thermal particle energy, $L_{\rm j}$ the jet luminosity, $n_{\rm s} (m,z)$ the (assumed stationary) stellar number density, $m$ the stellar mass, $z$ the jet height, $t$ the time, and $\frac{S_{\rm s} (m,t)}{S_{\rm j}}$ (or $\langle \frac{S_{\rm s} (m,t)}{S_{\rm j}} \rangle$) the (time average of the) fraction of jet area intercepted by one stellar interaction. 

One can integrate over the height of the jet the quantity:
\begin{equation}\label{eq:SigmaT}
\sigma_{\rm T} = \int \int \langle \frac{S_{\rm s} (m)}{S_{\rm j}} \rangle \ n_{\rm s} (m,z) \ {\rm d}m \ {\rm d}z\,.
\end{equation}
If the value of $\sigma_{\rm T}$ is much higher than $1$, it can be an indicator that the interaction is dynamically relevant for the jet, as all its section will be shaded by collisions with stars and their winds. In addition, $\sigma_{\rm T}\gg 1$ would mean that the jet-star collisions should take place in the wake of (many) other collisions further upstream of the jet.

When the jet interacts with a stellar wind, a double bow-shock is generated. The stagnation point is defined as the point where the wind and jet ram pressures are equal, and is located at a distance $R_{\rm s}$ from the star. This can define a section for the interaction with the jet, $S_{\rm s}=\pi R_{\rm s} ^2$. However, it has been shown using hydrodynamical simulations that kinetic energy is converted into internal energy at larger distances from the star. This implies that the dynamical interaction is effective significantly farther from the star than $R_{\rm s}$ with respect to kinetic energy dissipation, increasing the effective area of the shock by a factor $A=10-100$ \citep{bosch-ramon2015}. The pressures at the stagnation point for the stellar wind and for the jet are:
\begin{equation}
P_{\rm s} = \rho\,v^2_{\rm w} = \frac{\dot{M}\,v_{\rm w}}{4\,\pi\,R_{\rm s}^2} \ , \ \ \ \ P_{\rm j} \simeq 
\frac{L_{\rm j}}{c\, S_{\rm j}}\,,
\end{equation}
respectively, where $c$ is the speed of light, $\dot{M}$ the stellar mass-loss rate, and $v_{\rm w}$ the stellar wind speed. At the stagnation point, $P_{\rm s}=P_{\rm j}$, thus
\begin{equation}\label{eq:SsSj}
\dfrac{S_{\rm s} (m,t)}{S_{\rm j}}= \dfrac{A \ \pi \ R_{\rm s}^2 (m, t)}{\pi \ R_{\rm j}^2}=\dfrac{A \ c \ \dot{M}(m,t) \ v_{\rm w} (m,t)}{4 \ L_{\rm j}}\,. 
\end{equation}

Consequently, for $\sigma_{\rm T}<1$, the non-thermal luminosity injected into the jet depends on the stellar density, wind velocity and mass-loss rate, that is, it does not depend on the jet power. The stars with high momentum rates ($\dot{M}v_{\rm w}$) are the most relevant for the interaction. Therefore, we focus here on high-mass stars for high-SFR AGN galaxies, and post-main sequence low-mass stars for massive AGN host galaxies with an old stellar population; for simplicity, both groups are modeled as main sequence OB stars, and red giants at different stages of evolution, respectively. Therefore, particularly high mass-loss phases of stars (supergiant, Wolf-Rayet, luminous blue variable, asymptotic giant branch) are not considered as they would be relatively rare, despite their impact being possibly dominant should they interact with the jet not far from its base.

\subsection{OB stars in star-forming galaxies}

Massive star-forming galaxies, such as ultra-luminous and luminous infrared galaxies (ULIRGs and LIRGs), can have SFR of hundreds to a thousand solar masses per year \citep[e.g.,][]{SanMir1996}. Studies of nearby ULIRGs have shown that these galaxies tend to concentrate most of the star formation in inner circumnuclear disks, of a few hundreds of parsec in radius and approximately a hundred parsec in height \citep[e.g.,][]{MedGue2014}. In such disks, the SFR can be as high as a few hundred solar masses per year \citep[e.g.,][]{DowSol1998, TenBra2014}. 
We consider here that the stellar population interacting with the jet is composed of young OB stars, being formed at the high rates typical of U/LIRGs, and distributed homogeneously in a circumnuclear disk. 

\subsubsection{Stellar number density}

The number of stars being formed per unit of mass, time and volume ($V$) is $\phi (m,r,t)$, which actually does not depend on location (i.e., radius $r$ from the galaxy centre in spherical coordinates) for a homogeneous spatial distribution. Assuming that the SFR is constant in time \citep[see][]{araudo2013}, a homogeneous spatial distribution of stars within the disk, and a power-law dependence on the mass, $\phi (m,r,t)$ can be expressed as:
\begin{equation}
\phi = K \left( \dfrac{m}{M_\odot} \right)^{-x},
\end{equation}
where $x \sim 2.3$ in the $0.1 \leq m/M_\odot \leq 120$ range considered \citep{Sal1955,Kro2001}, and $K$ is a normalization constant with units $[K]=M_\odot^{-1}$ yr$^{-1}$ pc$^{-3}$. The star formation rate is $\dot{M}_{\rm SFR}=\iint \phi \ m \ {\rm d}m \ $d$V$:
\begin{equation}
\dot{M}_{\rm SFR}=K \pi R_\mathrm{d}^{2} \ h_\mathrm{d} \int_{0.1\,M_\odot}^{120\,M_\odot} \left( \dfrac{m}{M_\odot} \right)^{-x+1} {\rm d}m\,,
\end{equation}
where $R_\mathrm{d}$ is the stellar disk radius and $h_\mathrm{d}$ the total disk thickness. Along with $\dot{M}_{\rm SFR}$, these quantities can be known for a given galaxy; thus, the constant $K$ can be obtained. 

As stars are being born, they accumulate in the galaxy. For stars of masses such that $t < t_{\rm life}$, where $t_{\rm life}$ is the stellar lifetime, the density of stars is 
\begin{equation}
n_{\rm s}(m) = \int_0^t \phi(t',m) \mathrm{d} t' \approx \phi(t=0) \cdot t\,.
\end{equation}
For $t>t_{\rm life}$, the massive stars have started to die, and the distribution becomes steeper than $-2.3$. Then, the stellar density becomes $n_{\rm s}(m)=\phi(m,t=0)\cdot t_{\rm life}(m)$, with   
\begin{equation}\label{eq:tlife}
t_{\rm life}(m)= 10^a \left( \dfrac{m}{M_\odot} \right)^{-b}\,{\rm yr}.
\end{equation}
We consider $a=9.9,\ b=2.9$ in the range $1.25 \leq m/M_\odot \leq 3$; $a=9.6,\ b=2.4$ in $3 \leq m/M_\odot \leq 7$; $a=9.1,\ b=1.8$ in $7 \leq m/M_\odot \leq 15$; $a=8.0,\ b=0.8$ in $15 \leq m/M_\odot \leq 60$; and $t_{\rm life}\approx 0.004$~Gyr at $m> 60\,M_\odot$ \citep{EksGeo2012}.

\subsubsection{Mass-loss rate and wind speed}

To estimate $S_{\rm s}(m)$, assumed constant in time for an OB main sequence star, it is necessary to know $\dot{M}$ and $v_{\rm w}$. We follow the prescriptions in \citet{VinKot2000} derived for OB stars. For O stars ($16 \leq m/M_\odot \leq 120$),
\begin{equation}\label{eq:vinkO}
\begin{aligned}
\log \dot{M}(m) = &-6.7 + 2.2\ \log(L_{\rm s} /10^5\,L_\odot) -1.3\ \log(m/30\,M_\odot) \\ &-1.2\ \log(\frac{v_{\rm w}/v_{\rm esc}}{2}) +0.9\ \log(T_{\rm eff}/40000\,{\rm K}) \\ &- 10.9\ [\log(T_{\rm eff}/40000\,{\rm K})]^2 + 0.85\ \log(Z/Z_\odot),
\end{aligned}
\end{equation}
where $L_{\rm s}$, $T_{\rm eff}$ , and $Z$ are the luminosity, effective temperature, and metallicity of the star, respectively. The terminal wind velocity of the stars in this range is $v_{\rm w}\approx 2.6\ v_{\rm esc}$.

For B stars ($2 \lesssim m/M_\odot \leq 16$),
\begin{equation}\label{eq:vinkB}
\begin{aligned}
\log \dot{M}(m) = & -6.7 + 2.2\ \log(L_{\rm s} /10^5\,L_\odot) -1.3\ \log(m/30\,M_\odot) \\ &-1.6\ \log(\frac{v_{\rm w}/v_{\rm esc}}{2}) +1.1\ \log(T_{\rm eff}/20000\,{\rm K}) \\ & +0.85\ \log(Z/Z_\odot).
\end{aligned}
\end{equation}
The terminal wind velocity of the stars in this range is $v_{\rm w} \approx 1.3\ v_{\rm esc}$ for $T_{\rm eff}> 12 500$ K, and it drops to $v_{\rm w} \approx 0.7 \ v_{\rm esc}$ for $T_{\rm eff}< 12 500$ K \citep{LamSno1995}. 

Simple dependencies of the parameters with the stellar mass are assumed: $L_{\rm s} \propto m^{3.5}$ in the $2 \leq m/M_\odot \leq 50$ range, $L_{\rm s} \propto m$ in the $50 \leq m/M_\odot \leq 120$ range, $R_{\rm s} \propto m^{0.6}$, and $T_{\rm eff}=(\frac{L_{\rm s}}{4 \pi \sigma R_{\rm s}^2})^{1/4}$.

Metallicity was measured by \citet{HuoXia2004} in the central regions of some nearby ULIRGs, among them one of the objects studied in this work, Mrk 231, obtaining values of $Z \simeq Z_{\odot}$. We assume here solar metallicity for ULIRG-type galaxies, which implies that the last term in Eqs.~(\ref{eq:vinkO}) and (\ref{eq:vinkB}) does not contribute to the mass loss of their stars. It would have a significant impact, however, when deriving mass-loss rates for massive stars in environments significantly more metal-poor than our own galaxy (e.g., ULIRGs at $z \sim 2-3$).

The prescriptions given above do not account for two known discrepancies between theoretical and observational mass-loss rates: clumping, and the weak-wind problem.

Wind-clumping refers to density inhomogeneities in the stellar wind, and not considering them causes an overestimation of the mass-loss rates that can amount to factors of 2 to 10, depending on the specific diagnostics used to derive the observational values \citep{PulVin2008}. Analytical models need to be corrected by the square root of the Clumping factor ($\dot{M_{\rm cl}}=\dot{M} \cdot f_{\rm cl}^{-1/2}$) before being adjusted to observational data. Comparisons with the \citet{VinKot2000} model, which does not account for clumping, find discrepancies between the theoretical model and empirically derived mass-loss rates of a factor 2-3 lower \citep[e.g.,][]{SurHam2013,SunPul2011,Smi2014}. \citet{PulVin2008} suggest a maximum correction for theoretical models of a factor of 2. In order to be conservative, we reduce the mass-loss rate values given by Eqs.~(\ref{eq:vinkO}) and (\ref{eq:vinkB}) by a factor of 3. 

The weak-wind problem refers to the fact that empirically derived mass-loss rates for late O-/early B-type stars might be a factor $10-100$ lower than theoretically expected. The first statistically relevant evidence was provided by \citet{ChlGar1991}, and was confirmed by many later studies using UV line diagnostics \citep[see][and references therein]{PulVin2008}. However, later results show that the weak-wind problem is reduced or eliminated when taking into account a hotter component of the wind, as the wind is not weak, but its bulk is only detectable in X-rays \citep[e.g.,][]{HueOsk2012}. Still, a reduction of a factor of $3$ in the mass-loss rates of massive O-types to account for clumping, and a reduction of a factor of $10$ for late O-/early B-type stars to account for both clumping and weak winds, is suggested in a review by \citet{Smi2014}. 

\begin{figure}[!ht]
\centering
%\hfill
\includegraphics[width=0.50\textwidth,keepaspectratio]{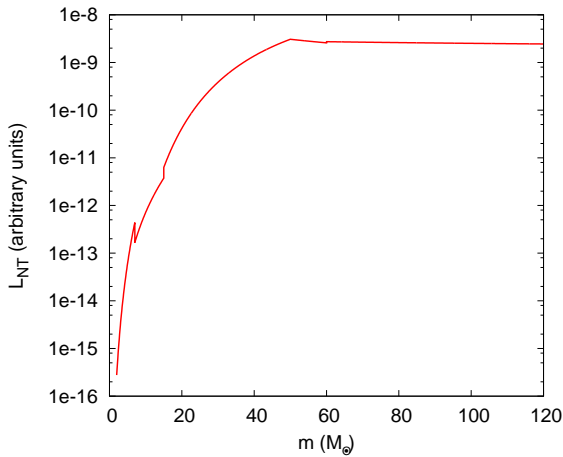}
\caption{Contribution to the non-thermal luminosity injected into the jet through stellar interactions for different stellar masses.}
\label{fig:OB_massloss}
\end{figure}

When applying Eqs.~(\ref{eq:vinkO}) and (\ref{eq:vinkB}), we correct by a factor of $3$ for clumping, and leave the weak-wind problem uncorrected due to the still unknown optimal reduction factor. However, as seen in Fig.~\ref{fig:OB_massloss}, where a quantity $\propto m\times dL_{\rm NT}/dm$ is shown, stars with masses significantly below $40\,M_\odot$ do not contribute significantly to the non-thermal luminosity; and the weak-wind problem would start to be significant for stars of spectral type O7-O8, which have masses of $\sim 25 - 28\,M_\odot$ \citep{Smi2014}. Therefore, for this study, correcting for weak winds becomes unnecessary. 

Following the given prescriptions and assumed dependencies, the mass-loss rate depends only on the mass. One can compute a weighted average over the mass that is independent of the SFR or the physical characteristics of a particular galaxy. This is valid as long as enough stars of a given mass are present to be treated as a continuum distribution to the needed degree of accuracy. The average mass-loss rate is
\begin{equation}
\langle \dot{M} \rangle = \dfrac{\int \dot{M}(m)n_{\rm s}(m){\rm d}m}{\int n_{\rm s}(m) {\rm d}m} = 3.4 \times 10^{-9}\,M_\odot\,\mathrm{yr}^{-1},
\end{equation}
in the considered $2 \leq m/M_\odot \leq 120$ range. Following the same procedure, a mass-averaged wind momentum can be derived: $\langle \dot{M}v_{\rm w} \rangle= 9.0 \times 10^{25}$~g~cm~s$^{-2}$.

The total number of stars with masses $\gtrsim 40\,M_\odot$ within the jet, for any given galaxy is
\begin{equation}
N_{\rm OB} = 13 \left( \dfrac{\dot{M}_{\rm SFR}}{100\,M_\odot\,{\rm yr} ^{-1}} \right) \left( \dfrac{h_{\rm d}}{100 \ {\rm pc}} \right)^2 \left( \dfrac{300\ {\rm pc}}{R_{\rm d}} \right)^2 \left( \dfrac{\theta}{0.1} \right)^2,
\end{equation}
where $\theta$ is the jet opening angle. Despite the fact that we adopt the continuum distribution assumption here, this result shows that it is only marginally valid.

\subsection{Red giants in elliptical galaxies}\label{elliptical}

Elliptical galaxies have in general very low SFRs, and therefore do not have a significant population of young OB stars that can interact with the jet. However, red giants can have high mass-loss rates, in the range of $\sim 10^{-10} - 10^{-5}\,M_\odot$~yr$^{-1}$ \citep{Rei1975}, and are abundant in this type of galaxies.

In the characterization of elliptical galaxies we assume the red giants to be distributed in an inner spherical bulge, with a density that decays as a power-law with the radial distance from the galaxy center. We also assume that there is no on-going star formation. 

\subsubsection{Stellar number density}

Knowing the mass profile of any particular elliptical galaxy, we can estimate the total mass of stars contained inside the bulge ($M_{\rm T}$). Then we use the Salpeter initial mass function (IMF) and normalize it to the total mass of stars:
\begin{equation}\label{eq:RedGiant_MT}
M_{\rm T}=K \dfrac{4}{3}\pi R_{\rm b}^{3} \int_{0.1\,M_\odot}^{m_2} \left( \dfrac{m}{M_\odot} \right)^{-x+1} \mathrm{d}m\,,
\end{equation}
with $x=-2.3$, and where $R_{\rm b}$ is the radius of the spherical bulge, and $m_2$ the mass of the stars in the galaxy exiting the red giant phase in the lifetime of the galaxy, that is, the largest stellar mass available. 

Knowing the lifetime of a galaxy ($t_{\rm gal}$), the red giants in the galaxy are those with masses between $m_1$ and $m_2$, with $m_1=m(t_{\rm life}=t_{\rm gal})$ being the mass of the stars entering the red giant phase at a time equal to the age of the galaxy, and $m_2=m(t_{\rm life}=t_{\rm gal}-t_{\rm rg})$ being the mass of the stars which entered the red giant phase exactly one red-giant lifetime before. We assume the lifetime of a red giant to be $\sim 5\%$ of the main sequence lifetime, and thus fix $t_{\rm rg}=0.05\,t_{\rm gal}$. We note that this approach assumes that all stars have been formed a time $t_{\rm gal}$ ago, that is star formation extended in time is not considered. Some stars may have formed later, which would enter the red giant phase at $t_{\rm gal}$ with higher masses, and would then lose more mass in the red giant phase. 

Since the lifetime of a galaxy is much larger than the lifetime of a red giant, $m_1$ and $m_2$ will be very similar. For a lifetime similar to that of the Milky Way, these masses are $\approx 0.83 \, M_\odot$. Then, we obtain the total number of red giants as:
\begin{equation}\label{eq:RedGiant_NT}
N_{\rm T}=K \dfrac{4}{3}\pi R_{\rm b}^{3} \int_{m1}^{m_2} \left( \dfrac{m}{M_\odot} \right)^{-x} \mathrm{d}m\,.
\end{equation}

Since all the red giants have very similar masses, and the total number of them is given by Eq.~(\ref{eq:RedGiant_NT}), together with the fact that their mass-loss rate and wind velocities are mass-independent (see section \ref{RGmassloss+wind}), there is no need to maintain a mass dependency on the number density. However, in this case, since we are considering a large and spherical bulge, there is a decay of the density with radial distance/jet height ($z$), that is, $n_{\rm s}(z)\propto N_{\rm T}/z^\xi$. We assume this dependence to be a power law, and consider two values for the index $\xi$: $\xi=1$, which is the stellar index estimated for M87 \citep[derived from][]{gebhardt2009}, and $\xi=2$ for comparison.

\subsubsection{Mass-loss rate and wind speed} \label{RGmassloss+wind}

The mass-loss rate of a red giant depends on its luminosity and radius, following:
\begin{equation}\label{eq:MdotRG}
\dot{M}=4 \times 10^{-13} \left(\frac{L}{L_\odot}\right)\left(\frac{g_\odot}{g}\right)\left(\frac{R_\odot}{R}\right)\,M_\odot\,{\rm yr}^{-1},
\end{equation}
where $g$ is the stellar surface gravity \citep{Rei1975}.

As a red giant evolves, more hydrogen from the $H$-burning shell surrounding the core turns into helium, increasing the mass of the {\sl He}-core, and the stellar radius and luminosity. Therefore, the mass-loss rate has a time dependence for a red giant star.

\cite{Jos1987} provide a fit, based on numerical models, to the core mass-luminosity relation for red giants with core masses in the range of $0.17\,M_\odot \lesssim m_c \lesssim 1.4\,M_\odot$:
\begin{equation}\label{eq:Lmc}
\begin{aligned}
L(m_{\rm c}) \simeq \dfrac{10^{5.3}\mu^{6}}{1+10^{0.4}\mu^{4}+10^{0.5}\mu^{5}}\,L_\odot \,, &&\mu \equiv \dfrac{m_{\rm c}}{M_\odot}\,,
\end{aligned}
\end{equation}
and a fit to the core mass-radius relation in the same range:
\begin{equation}\label{eq:Rmc}
R(m_{\rm c}) \simeq \dfrac{3.7 \times 10^3 \mu^4}{1+ \mu^3 1.75 \mu^4} R_\odot\,.
\end{equation}

As all the red giant stars in the galaxy have a very similar mass, we consider them all to have the same initial core mass, and consequently the exact same mass-loss rate as a function of time. The time dependence can be introduced when considering that the dominant energy source in red giants is the p$-$p chain, with a $\sim 0.7$~\% efficiency, and the $He$-core mass increases as hydrogen burns into helium according to \citep{SyeUlm1999}:

\begin{equation}
L(m_{\rm c}) \simeq 0.007\,M_\odot c^2 \dot{\mu}\,.
\end{equation}

The core-mass range considered in this work is $0.17 - 0.43\,M_\odot$: starting with an initial core mass that corresponds to the lower limit of the range for which Eqs.~(\ref{eq:Lmc}) and (\ref{eq:Rmc}) are fitted, and stopping at a value for which the radius, mass-loss and lifetime of the red giant are reasonable ($R_{\rm f}\sim 110\ R_\odot$, $\dot{M}_{\rm f} \sim 5 \times 10^{-8} \, M_\odot$~yr$^{-1}$ and $t_{\rm rg} \sim 7.3 \times 10^8$ yr). These values are limited by the considered initial mass of the star, as the sum of the final core mass and total mass lost cannot exceed it. Unlike with OB stars, Eq.~(\ref{eq:MdotRG}) does not include a metallicity dependence. Red giants are assumed here to lose along their lifetime all the mass that does not go into the core, independently of metallicity, and therefore the final average mass-loss rate is the same regardless of the specific effects of metallicity on the $\dot{M}(t)$-curve.

Wind speeds of red giants are relatively low, typically $\lesssim 10^7$~cm~s$^{-1}$ \citep[e.g.,][]{CroPhD2006, EspCro2008}. In this work we take, for simplicity, $v_{\rm w} = 10^7$ cm s$^{-1}$, considering it constant during the star evolution.

\begin{figure}[!ht]
\centering
%\hfill
\includegraphics[width=0.50\textwidth,keepaspectratio]{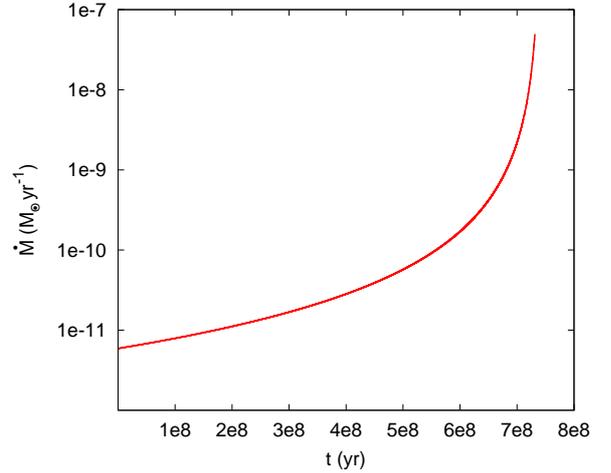}
\caption{Mass-loss rate as a function of time for a red giant with a core mass ranging from $0.17\, M_\odot$ to $0.43\,M_\odot$, in a lifetime of $\sim 7.3 \times 10^8$ yr.}
\label{fig:RG_massloss}
\end{figure}

As seen in Fig.~\ref{fig:RG_massloss}, the latter stages of the life of a red giant ($\sim 0.01 \ t_{\rm rg}$, which means, $\sim 1 \ \%$ of red giants within the jet) contribute most significantly to the mass-loss rate and, therefore, to the injected non-thermal luminosity. For Eqs.~(\ref{eq:NTLum}) and (\ref{eq:SsSj}), as the mass-loss rate and wind speed are mass-independent, they can both be used as constants if $\langle \dot{M} \rangle$ is time-averaged:
\begin{equation}
\langle \dot{M} \rangle = \dfrac{\int \dot{M}(t) {\rm d}t}{t_{\rm rg}} = 5.7 \times 10^{-10}\,M_\odot \mathrm{yr}^{-1}.
\end{equation}
Then, taking the considered constant value for the wind speed, we obtain an average wind momentum of $\langle \dot{M}v_{\rm w} \rangle= 3.6 \times 10^{23}$~g~cm~s$^{-2}$.

\section{Jet model}\label{model}

We adopt a jet with a conical geometry, that is, a constant opening angle $\theta$, which is launched at a distance $z_{0}$ above the supermassive black hole in the center of the galaxy. The radius of the jet is a function of the distance $z$ to the black hole:
\begin{equation}\label{eq:rJet}
R_{\rm{j}}(z) = \theta\,z\,.
\end{equation}
Assuming equipartition between the magnetic field and the jet total energy density, the magnetic field in the jet base would be
\begin{equation}
\frac{B^{2}_0}{8 \pi} =  \frac{1}{2}\frac{L_{\rm{j}}}{\pi R_{\rm{j}}(z_{0})^2 c}\,,
\end{equation}
where $B_0 = B(z_0)$. The magnetic field decreases with $z$ according to 
\begin{equation}
B(z) = B_0 \left(\frac{z_0}{z}\right)^{m},
\end{equation}
with $1 \leq m \leq 2$, depending on the topology of the magnetic field.

Polarization angles from blazars are found to be either nearly transverse or nearly parallel to the jet axis; this dualism is consistent with magnetic fields that are intrinsically oblique, but the observed directions are altered by relativistic effects \citep{marscher2002}. For toroidal and poloidal fields comparable in the jet frame, since $B_{\phi}/B_{\rm z} \geq \Gamma$, a relativistic jet would be dominated by the toroidal magnetic field in the observer frame \citep{lyutikov2004}.

In this work, we parametrize the magnetic pressure through a fraction $\zeta_{\rm eq}$ of the equipartition value. Also, in most cases the magnetic field is assumed to be predominantly perpendicular to the flow motion, so we adopt $m=1$ \citep{spruit2010}. Therefore, in the flow frame one obtains
\begin{equation}\label{eq:Bfield}
B'_{\phi}(z) = \frac{1}{\Gamma z}\sqrt{\frac{4\zeta_{\rm eq}L_{\rm{j}}}{\theta^2c}}\,,
\end{equation}
where $\Gamma$ is the jet bulk Lorentz factor, which is considered to be constant.
We consider two cases: $\zeta_{\rm eq}=1$, which means that $B(z)$ is in equipartition, and $\zeta_{\rm eq}=10^{-2}$ as an example of a magnetic field below equipartition. In the case where we consider a dominant poloidal field, $m=2$ (see Sect. \ref{results}; a poloidal magnetic field decays faster than a toroidal field with $z$), we take into account that it remains invariant between the observer and jet frames, that is $B_{\rm z}(z) = B'_{\rm z}(z)$.

\section{Galaxy hosts}\label{hosts}

The prescriptions described in Sect. \ref{stellar} and the jet model described in Sect. \ref{model} are applied to specific galaxy types to estimate the contribution of jet-star interactions to high-energy radiation in realistic contexts.

\subsection{Elliptical galaxy: M87}

As a fiducial elliptical galaxy, we consider the case of  M87. The galaxy bulge has a radius of $\sim 40''$ \citep{HarBir1999}, which corresponds to $R_{\rm b} \sim 3.1$ kpc. Knowing the bulge size, we estimate the total mass contained within it from \citet{gebhardt2009}, and determine the total number of red giants using  Eqs.~(\ref{eq:RedGiant_MT}) and (\ref{eq:RedGiant_NT}). The number of red giants within the bulge is $\approx 1.3 \times 10^9$.

The jet in M87 has a luminosity of $L_{\rm j} = 10^{44}$ erg s$^{-1}$ \citep{owen2000}, an inclination angle of $20^{\circ}$ \citep{acciari2009}, and an opening angle of $\sim~0.1$~rad \citep{BirMei1993,DoeFis2012}. Taking this aperture, the number of red giants within the jet would be $\sim 3.2 \times 10^6$. Radio lobes detected by \citet{owen2000} show that the emission in M87 comes from a region within $\sim 40$ kpc. The jet remains undisturbed and collimated only for a few kpc, where it is relativistic, with a Lorentz factor of $\Gamma\sim 2-3$ \citep{biretta1995}. Thus, here we focus on an extension of $z_{\rm max} = 5$~kpc and adopt a Lorentz factor of $\Gamma=3$.

\begin{figure}[!ht]
\centering
%\hfill
\includegraphics[width=0.50\textwidth,keepaspectratio]{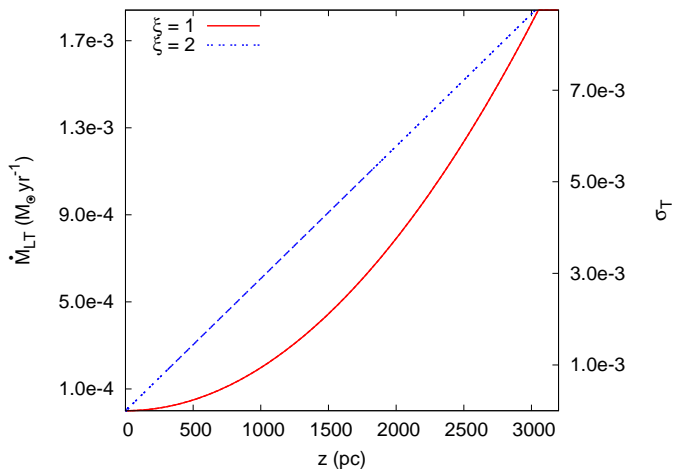}
\caption{Total mass loaded into the jet (left axis), and total fraction of jet surface intercepted by stellar interactions (right axis), as a function of jet height, for $\xi=1$ and 2, in M87.}
\label{fig:McM87}
\end{figure}

We plot in Fig. \ref{fig:McM87} the total loaded mass rate and the total surface of interaction as defined in Eq.~(\ref{eq:SigmaT}). At the total bulge height, we obtain ratios $\Gamma\,\dot{M} c^2 / L_j \sim 3$ and $\sigma_T \sim 0.01$, which mean that wind mass-load and subsequent jet slow down is likely important in the jet of M87 on kpc scales. On the other hand, only 1\% of the jet section is covered by interactions, which may mean that the loaded mass is confined only to relatively small regions of the jet. However, given the unstable nature of jet-wind interactions and the subsequent loaded matter evolution \citep[see, e.g.,][]{bosch-ramon2012,moreno2016}, plus the complex dynamic pattern arising from such an inhomogeneous configuration,  it seems more likely that the loaded wind material will effectively spread all over the jet. It is worth noting that both the number of stars within the jet and the jet mass-load estimates derived here are similar to those found by \cite{wykes13,wykes15} for the radiogalaxy Centaurus~A.

\subsection{Starburst galaxies: Mrk~231 and 3C~273}

We study two sets of parameters describing star-forming galaxies: One is considered to be a local universe galaxy with a weak jet and a very high SFR, for which we take the particular case of Mrk~231. The other starburst is the powerful quasar 3C 273.

Mrk~231:  The jet in Mrk~231 has a luminosity of $L_{\rm j} = 10^{43}$ erg s$^{-1}$ \citep{reynolds2009}. There is evidence supporting a jet viewed nearly along the line of sight, with an inclination 
$i <14^{\circ}$ and a high Lorentz factor \citep[e.g.,][]{reynolds2009, DavTac2004}; we adopt $i=10^{\circ}$, and a Lorentz factor of $\Gamma=6$. The size of the collimated radio source is estimated at $\sim 70$~pc \citep{taylor1999}, and its opening angle $\theta=0.1$. 

The stellar disk of Mrk~231 has a total thickness of $23$~pc and a radius of $\sim 300$~pc, and a nuclear SFR of $\sim 100 - 350\,M_\odot$~yr~$^{-1}$ \citep{DowSol1998,TenBra2014}. We take here the limit value of $350\,M_\odot$~yr~$^{-1}$. 

\begin{figure}[!ht]
\centering
%\hfill
\includegraphics[width=0.50\textwidth,keepaspectratio]{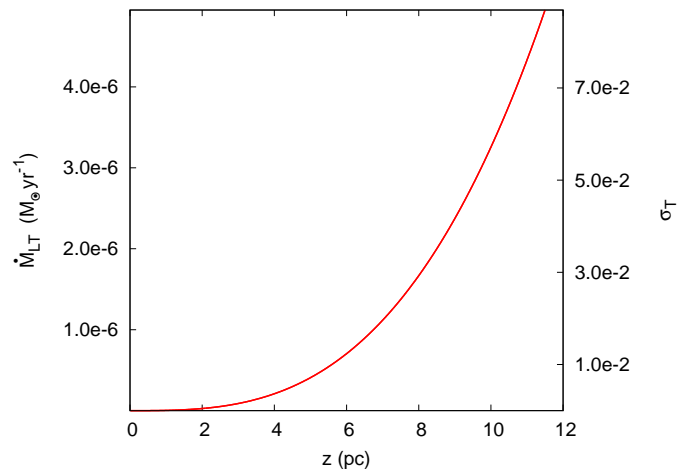}
\caption{Total mass rate loaded into the jet (left axis), and total fraction of jet surface intercepted by stellar interactions (right axis), as a function of jet height for Mrk~231.}
\label{fig:McM231}
\end{figure}

The total mass rate loaded inside the jet of Mrk~231 by stars and the total surface of interaction, as defined in Eq.~(\ref{eq:SigmaT}), are plotted in Fig.~\ref{fig:McM231} as a function of jet height. At the total jet height, we have ratios $\Gamma\,\dot{M} c^2 / L_j \approx 0.17$ and $\sigma_T \approx 0.09$. Unlike the case of M87, the jet of Mrk~231 seems to be only slightly mass-loaded and slowed down by the winds of massive stars, but in this case the loaded matter will spread inside the jet more efficiently due to a higher $\sigma_T$-value.

The source 3C~273, located at $z=0.158$, is one of the brightest and closest quasars. The jet luminosity is $L_{\rm j} \sim 10^{46}$~erg~s$^{-1}$ (e.g., \citealt{stawarz2004,ghisellini2010}; although lower intrinsic luminosities of $\sim 4\times 10^{44}$~erg~s$^{-1}$ were recently estimated by \citealt{punsly2016}). VLBI observations show a small-scale radio jet, whose components are characterized by apparent superluminal motions, indicating a jet close to the line of sight. In addition, radio observations also reveal a large-scale jet, that extends up to tenths of kpc \citep{conway1981}. The viewing angle of the larger jet, however, seems to differ from the one of the inner jet by $\sim 20^{\circ}$ \citep{stawarz2004}. Thus, we study the emission produced in the inner jet.

Superluminal motions are found up to a distance of hundreds of pc \citep{davis1991};  we consider that the jet extends up to a distance comparable to the stellar disk radius, that is $z_{\rm max} = R_{\rm d}$.

This source is highly variable at all wavelengths; a precessing inner jet \citep{abraham1999} and a double helix inside the jet \citep{lobanov2009} have been suggested as responsible for the radio variability. Since we do not attempt to model the multi-wavelength emission of the source nor its variability, we consider average values for the inclination angle and the Lorentz factor, adopting $i = 6^{\circ}$ and $\Gamma = 10$ respectively \citep{jorstad2005}. 

\begin{figure}[!ht]
\centering
%\hfill
\includegraphics[width=0.50\textwidth,keepaspectratio]{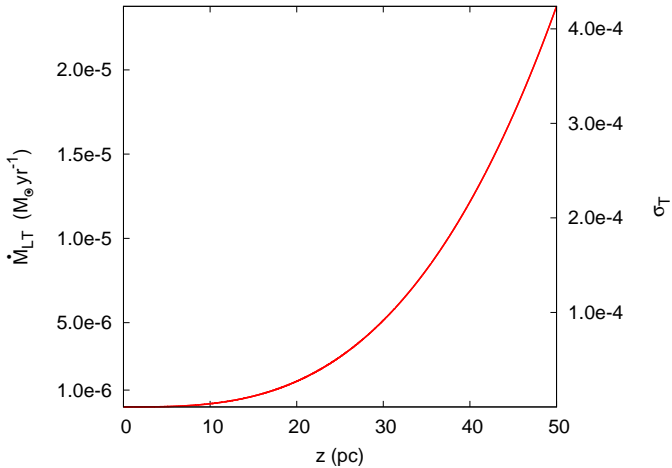}
\caption{Total mass rate loaded into the jet (left axis), and total fraction of jet surface intercepted by stellar interactions (right axis), as a function of jet height for 3C~273.}
\label{fig:McCygA}
\end{figure}

With an infrared luminosity of $\log(L_{IR}/L_{\odot})= 12.73$ \citep{KimSan1998}, 3C~273 is classified as a ULIRG. For the stellar disk properties, we take a total thickness of 100~pc (average value for nearby ULIRGs in \citealt{MedGue2014}), and a radius of 300~pc as in Mrk~231. SFR estimations for this object are $50 -150\ M_\odot$ yr$^{-1}$ \citep{FarLeb2013}, and $129\ M_\odot$ yr$^{-1}$ \citep{ZhaShi2016}, for the whole galaxy. As in this type of object, most of the star formation originates in the inner regions; we assume a SFR of $\sim 100\ M_\odot$ yr$^{-1}$ to be concentrated in the molecular disk. At the total jet height, we have ratios $\Gamma\,\dot{M} c^2 / L_j \sim 10^{-3}$ and $\sigma_T \sim 4 \times 10^{-4}$, as seen in Fig.~\ref{fig:McCygA}. Therefore, mass-load and the dynamical effects induced by stellar winds are likely minor in this source.

Table \ref{table1} lists all the relevant parameter values of the model and the sources.

\begin{table*}[ht]
    \caption[]{Main parameters of the model.}
        \label{table1}
        \centering
\begin{tabular}{lrrr}
\hline\hline %
Parameters & 3C~273  & Mrk~231 & M87 \\ [0.01cm]
\hline

$d$: distance [Mpc]    & $730$  & $180$  & $16$                                                         \\
SRF: star formation rate [$M_{\odot}$ yr$^{-1}$]    & $100$     & $350$  & - \\
$h_{\rm{d}}$: stellar disk thickness [pc]       & $100$ & $23$           & - \\
$R_{\rm{d}}$: stellar disk radius [pc]  & $300$ & $300$          & $3100^{\star}$ \\
$L_{\rm{c}}$: jet power [erg s$^{-1}$]   & $10^{46}$     & $10^{43}$     & $10^{44}$\\
$z_{\rm{max}}$: jet height [pc]  & $300$         & $70$  & $5000$\\
$\Gamma$: Lorentz factor                & $10$          & $6$           & $3$\\
$\theta$: opening angle [rad]  & $0.1$  & $0.1$ & $0.1$\\
$i$: inclination [$^{\circ}$] & $6$     & $10$  & $20$\\

\hline  \\[0.005cm]
\end{tabular}
\begin{tablenotes}
{\footnotesize
\item Notes: References for all parameter values are given in Sect. \ref{hosts}. Distances are taken from NED as of March 2017. $^{\star}$ Radius of the galactic bulge. The stars in M87 are assumed to be spherically distributed.\\}

\end{tablenotes}
\end{table*}

\section{Radiated non-thermal power}\label{appLnt}

Assuming that IC losses dominate, one can easily estimate the apparent luminosity of the high-energy emission expected from the interaction of a given AGN jet with the population of stars in the host galaxy. The radiative efficiency of the shocked fluid can be approximated as:
\begin{equation}\label{eq:frad}
f_{\rm rad}(E,z) = \frac{t^{-1}_{\rm rad}}{t^{-1}_{\rm rad} + t^{-1}_{\rm nrad}}\,,
\end{equation}
where $t^{-1}_{\rm nrad}$ accounts for the non-radiative losses (e.g., adiabatic losses, particle advection), and  $t^{-1}_{\rm rad}$ accounts only for IC losses in the Thomson regime\footnote{However, these losses could be associated as well to synchrotron losses under a magnetic field of equivalent energy density to the dominant photon field.}:
\begin{equation}
t^{-1}_{\rm IC,T} = \frac{4 c \sigma_{\rm T}}{3} \frac{E}{(m_ec^2)^2} \omega_{\rm ph}\,.
\end{equation}
The luminosity density generated only by the red giants in the galaxy is comparable to the one generated by the whole stellar population in the bulge \citep[derived from][]{gebhardt2009}. Here we consider as target photons for IC interactions those emitted by the whole red giant population. Given that other photon fields produced in the galaxy or its central region can be present, the radiative efficiency derived is rather conservative. The photon energy density at a given $z$-value is estimated as:
\begin{equation}\label{eq:wph}
\omega_{\rm ph} (z) = \int  \frac{ L_{\rm s}(m)n_{\rm s}(m,z) }{4\pi c (z^2+r^2-2rz\cos \theta)} dm dV\,,
\end{equation}
In the case of M87, $L_{\rm s}(m)$ should be replaced by $\left\langle L_{\rm s}\right\rangle$, the time average red giant luminosity.

The apparent non-thermal luminosity per unit volume at height $z$ due to jet-star interactions is then:
\begin{equation}\label{eq:appLnt}
\frac{{\rm d}L_{\rm NT}^{\rm app}(z)}{{\rm d}V} = \eta_{\rm NT} L_{\rm j}  f_{\rm rad}(z)  \frac{\delta_{\rm j}^4}{\Gamma_{\rm j}^2} \int{  \left\langle \frac{S_{\rm s}(m,z)}{S_{\rm j}(z)}\right\rangle n_{\rm s}(z,m) {\rm d}m}\,,
\end{equation}
where $n_{\rm s}(z,m)$ is the density of stars, and $\delta_{\rm{j}}$ is the Doppler boosting factor, given by
\begin{equation}\label{eq:doppler}
\delta_{\rm{j}} = \frac{1}{ \Gamma(1-\beta_{\rm j} \cos i )}\,,
\end{equation}
where $i$ is the inclination, that is, the angle between the jet axis and the line of sight.  Notice that Eq.~(\ref{eq:wph}) is the photon energy density in the laboratory frame; in the jet frame, it is enhanced by a factor $\sim \Gamma_{\rm j}^2$ (whereas the IC target photon energy is enhanced by $\sim \Gamma_{\rm j}$). We note, however, that $f_{\rm rad}$ is an invariant quantity. As mentioned in Sect.~\ref{stellar}, the effective area of the shock is larger than the one defined by the stagnation distance; we adopt here $A=100$ \citep{bosch-ramon2015}. The total radiative output is computed integrating Eq.~(\ref{eq:appLnt}) over the jet volume. 

We estimate the apparent non-thermal radiative output at a reference energy of $E'_{\rm IC} = (m_e c^2)^2/kT_{\rm s}\Gamma$, where $E'_{\rm IC}$ is approximately the maximum of the IC cross-section in the flow frame around the Thomson-Klein-Nishina (KN) transition. In the case of M87, we obtain  $L^{\rm app}_{\rm NT} \approx 5 \times 10^{-3} \eta_{\rm NT} L_{\rm j}$ at $E'_{\rm c} \approx 250$ GeV, for both values of the index of the stellar density. For Mrk~231, the apparent non-thermal luminosity is $L^{\rm app}_{\rm NT} \approx 7\times 10^{-2} \eta_{\rm NT} L_{\rm j}$ at $E'_{\rm IC} \approx 10$ GeV, whereas for 3C~273, it is lower than the jet luminosity, $L_{\rm NT}\approx 7\times 10^{-4} \eta_{\rm NT} L_{\rm j}$, with the same $E'_{\rm IC}$.  In Sect.~\ref{discussion} we discuss how reliable these estimates are.

\section{Non-thermal processes}\label{SED}

As shown in Sect.~\ref{appLnt}, $L^{\rm app}_{\rm NT}$ can easily reach $\sim 1$\% of the jet luminosity. In this section we study the non-thermal processes in more detail, and compute the synchrotron and IC spectral energy distributions (SEDs).

\subsection{Energy losses}\label{enlo}

We consider that particles lose energy by synchrotron radiation, IC interactions, and adiabatic cooling. We calculate the cooling rates in the flow frame. The cooling rate for synchrotron radiation is given by:
\begin{equation}\label{eq:coolsyn}
t'^{-1}_{\rm{synchr}}(E',z) = \frac{4}{3} \frac{c\sigma_T}{(m_ec^2)^2} \frac{B'^2(z)}{8\pi}E',
\end{equation}  
and adiabatic losses can be estimated as:
\begin{equation}\label{eq:coolad}
t'^{-1}_{\rm{ad}}(E',z) = \frac{2}{3} \frac{\Gamma c}{z}\,.
\end{equation} 

There are several radiation fields that can provide targets to IC interactions: Locally produced radiation, as synchrotron emission (synchrotron self-Compton, SSC), and external photon fields, such as the radiation from the stars in the galaxy, infrared (IR) photons from dense regions, or the cosmic microwave background (CMB).

In particular, IR photons have been considered in the starburst galaxies. The IR luminosity of the star-forming disk in Mrk~231 is estimated in $L_{\rm IR} = 3 \times 10^{12}\,L_{\odot}$~erg~s$^{-1}$; for the ULIRG 3C~273, we consider an IR luminosity of $L_{\rm IR} = 5.4 \times 10^{12}\,L_{\odot}$~erg~s$^{-1}$ \citep[both from][]{sanders1988}. We model the IR fields as gray bodies with a temperature of $\sim 200$~K. In the case of M87, the starlight and CMB photons are the most relevant targets  \citep{hardcastle2011}; thus, we include CMB photons for this source. The extragalactic background light (EBL) energy density is, however, at least a factor $30$ below the CMB \citep{cooray2016}; we therefore do not consider the EBL as an additional target. 

In addition, 3C~273 shows an excess in the optical/UV emission, likely the result of an accretion disk or reprocessing of radiation from a hot corona. The coronal emission is also observed on X-rays at $E \lesssim 30$ keV \citep{madsen2015}. The size of the accretion flow is estimated in $0.02-0.05$ pc \citep{chidiac2016}. As a result, its photon energy density is deboosted when seen from the jet. Moreover, the scattering probability for IC interactions is reduced  a factor $(1-\beta \cos \theta)^2 \sim 10^{-16}$. We consider, then, that this component does not provide a relevant target for our study.

The energy density of synchrotron photons is $10^{-4}-10^{-3}$ times lower than the magnetic energy density in $z_{\rm max}$ for both values of the magnetic field, for the three sources. Then, SSC have turned out to be irrelevant in our scenario, as expected given the large scales involved. In the timescale analysis, we focus on scales $\sim z_{\rm max}$ as in most of the cases explored here the largest scales are radiatively dominant.

The maximum energy that electrons can attain depends on the energy loss/gain balance. The acceleration rate is assumed to be: 
\begin{equation}\label{eq:accrate}
t'^{-1}_{\rm{acc}}(E',z) = \eta \frac{ecB'(z)}{E'}\,,
\end{equation} 
where $\eta = (v/c)^2 /2\pi$ and it approaches $0.1$ as $v \rightarrow c$ \citep[we follow the same approach as in][further details are provided there]{moreno2016}. 

\begin{figure*}[!ht]
\centering
%\hfill
\subfigure[3C~273: $z= 300$ pc.]{\label{fig:perdidas:a}\includegraphics[width=0.45\textwidth,keepaspectratio]{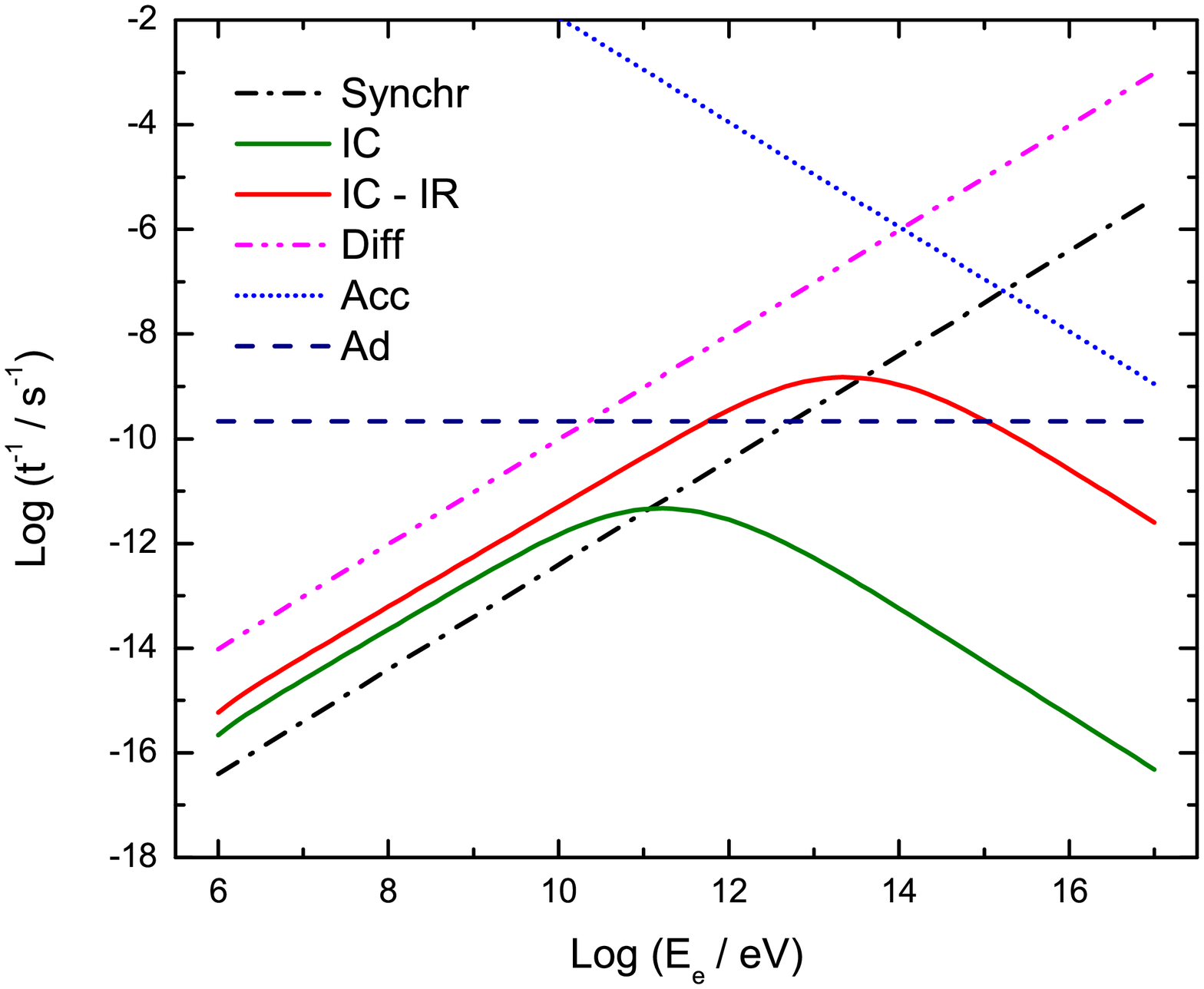}} \hspace{20pt} 
\subfigure[Mrk~231: $z= 70$ pc.]{\label{fig:perdidas:b}\includegraphics[width=0.45\textwidth,keepaspectratio]{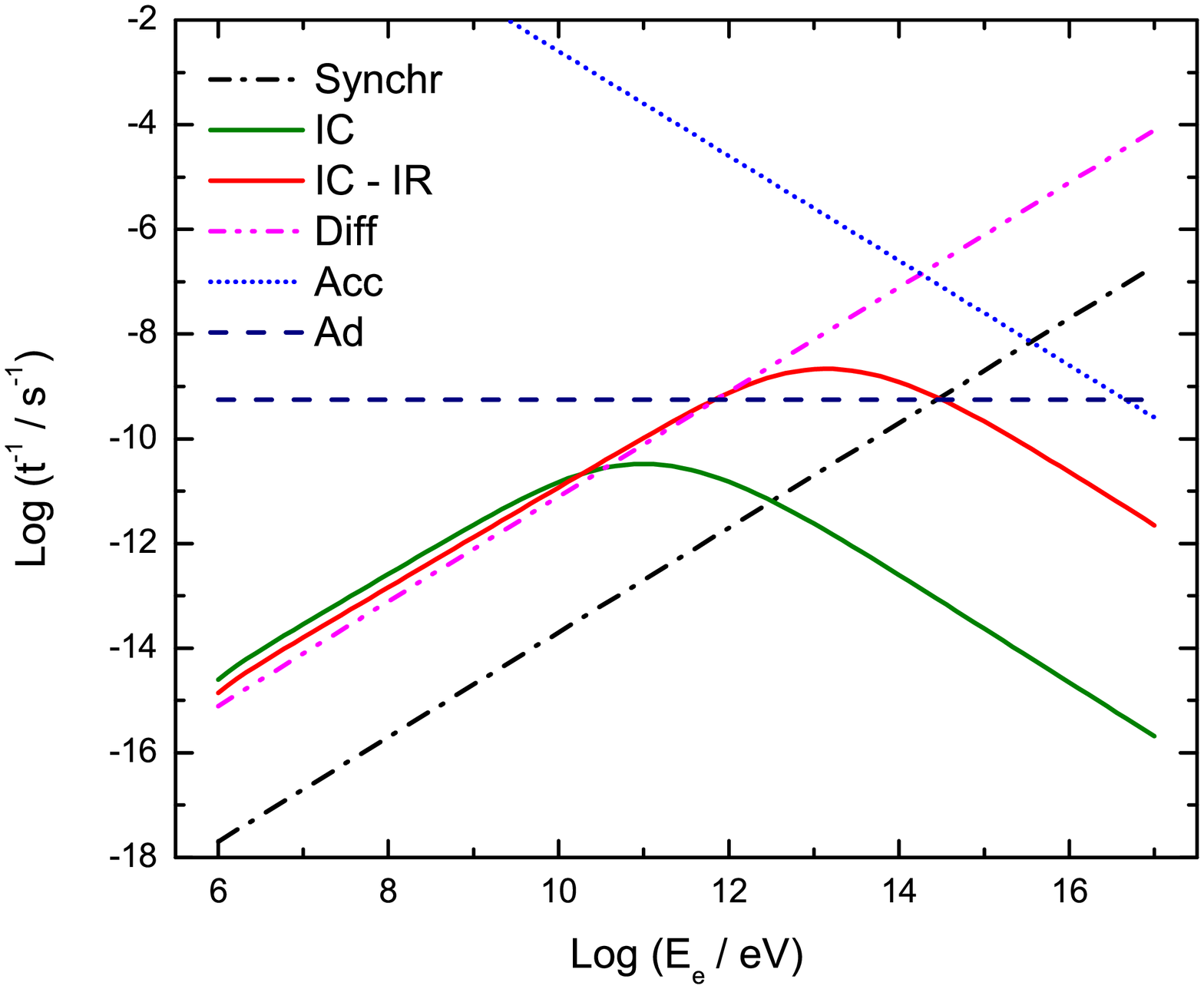}}\hspace{20pt} 
\caption{Energy losses at $z_{\rm max}$ for the star-forming galaxies in the low-$B$ case.}
\label{fig:losses_starburst}
\end{figure*}

\begin{figure*}[!ht]
\centering
%\hfill
\subfigure[$B=B_{\rm eq}$]{\label{fig:perdidas:am87}\includegraphics[width=0.45\textwidth,keepaspectratio]{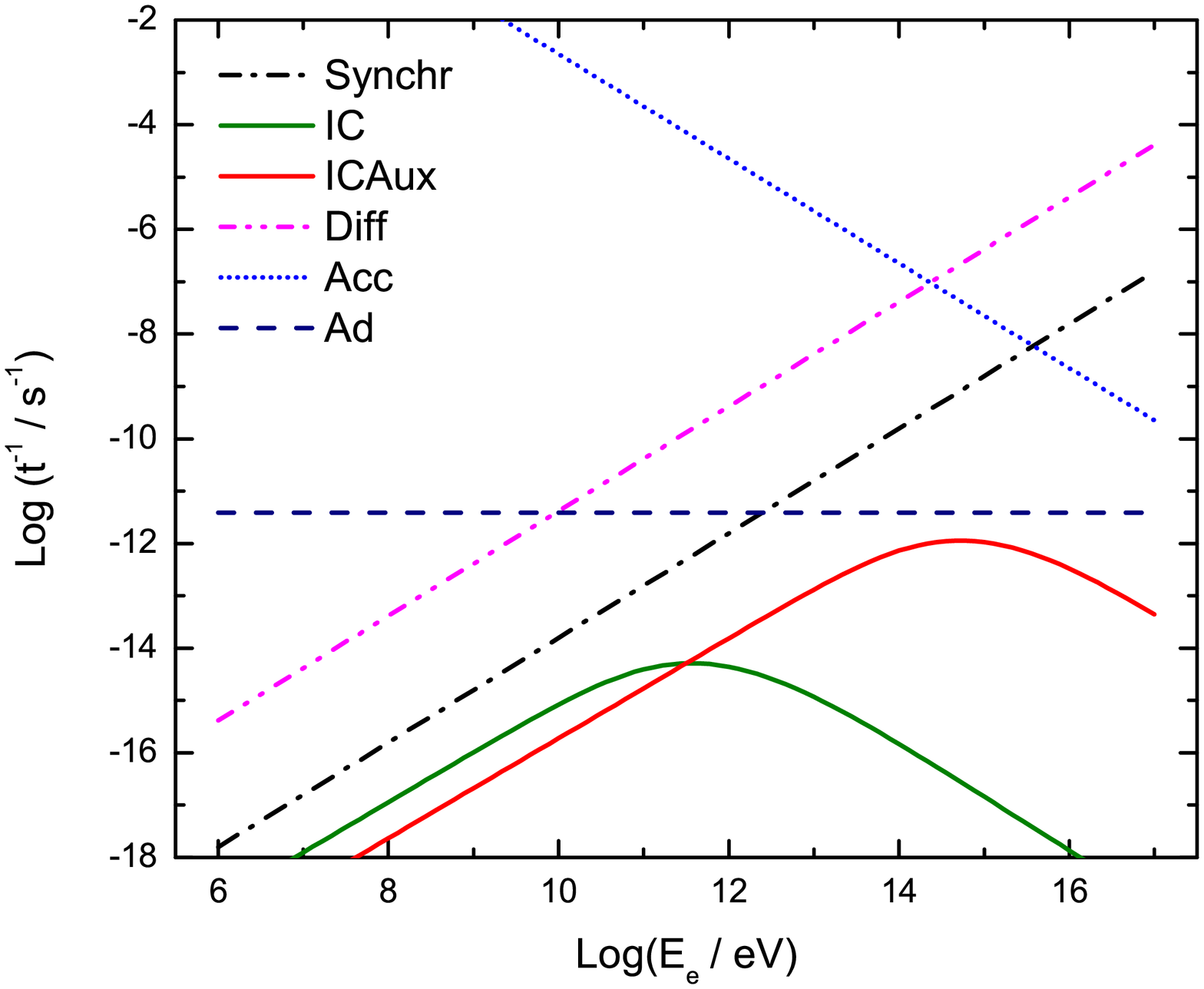}} \hspace{20pt} 
\subfigure[$B=10^{-1}B_{\rm eq}$]{\label{fig:perdidas:bm87}\includegraphics[width=0.45\textwidth,keepaspectratio]{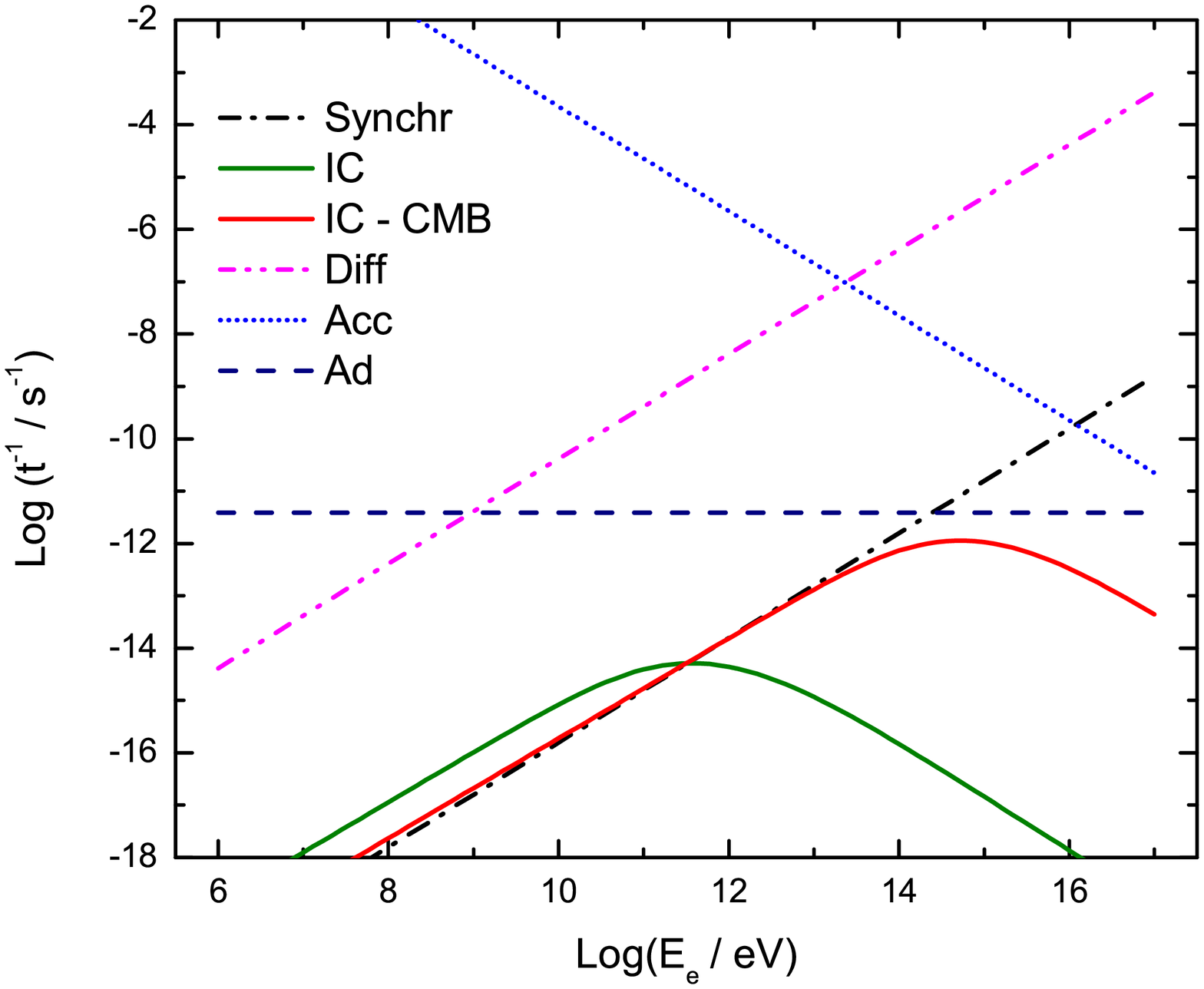}}\hspace{20pt} 
\caption{Energy losses at $z_{\rm max} =5$~kpc for M87, in the case $\xi=1$. }
\label{fig:losses_M87}
\end{figure*}

Figure~\ref{fig:losses_starburst} shows the cooling rate at $z_{\rm max}$, together with the acceleration rate for the starburst galaxies, for the sub-equipartition value of the magnetic field. Figure \ref{fig:losses_M87} shows the cooling rate at $z_{\rm max}$ for M87, both for $B=B_{\rm eq}$ and $B=10^{-1} B_{\rm eq}$, in the case of the stellar index $\xi=1$. Similar results (not shown here) to those presented in Fig.~\ref{fig:losses_M87} are obtained when considering the stellar index $\xi=2$. 

In all the explored cases, the maximum electron energy is determined either by synchrotron losses or diffusion out of the accelerator (i.e., the jet-star direct interaction region).

\subsection{Particle injection}

We adopt an injection function, in units of particles per time and energy unit for a jet height interval ${\rm d}z$, given by
\begin{equation}
\rm{d}Q(E',z) = Q_0(z) E'^{-\alpha} \exp(-E'/E'_{\rm max}(z))\,,
\end{equation}
where the injection index is taken as $\alpha=2$, characteristic of diffusive acceleration mechanisms. The normalization function $Q_0(z)$ depends on the available non-thermal energy, quantified by Eq.~(\ref{eq:NTLum}). Since the particle energy in the flow frame is $E'_{\rm NT} = E_{\rm NT} / \Gamma$, and the cell crossing time in the flow frame is $\Gamma$ times longer, d$Q(E',z)$ can be normalized through
\begin{equation}
\int^{E'_{\rm max}}_{E_{\rm min}} {{\rm d}E' E' \rm{d}Q(E',z)}\approx\frac{1}{\Gamma^2}{\rm d}L_{\rm NT}(z) \,,
\end{equation}
where ${\rm d}L_{\rm NT}(z)$ is the non-thermal injected luminosity within the jet height interval ${\rm d}z$. The transport equation of electrons is solved following the approach described in \citet{moreno2016}.

\begin{figure*}[!ht]
\centering
\includegraphics[width=0.3\textwidth,angle=270,keepaspectratio]{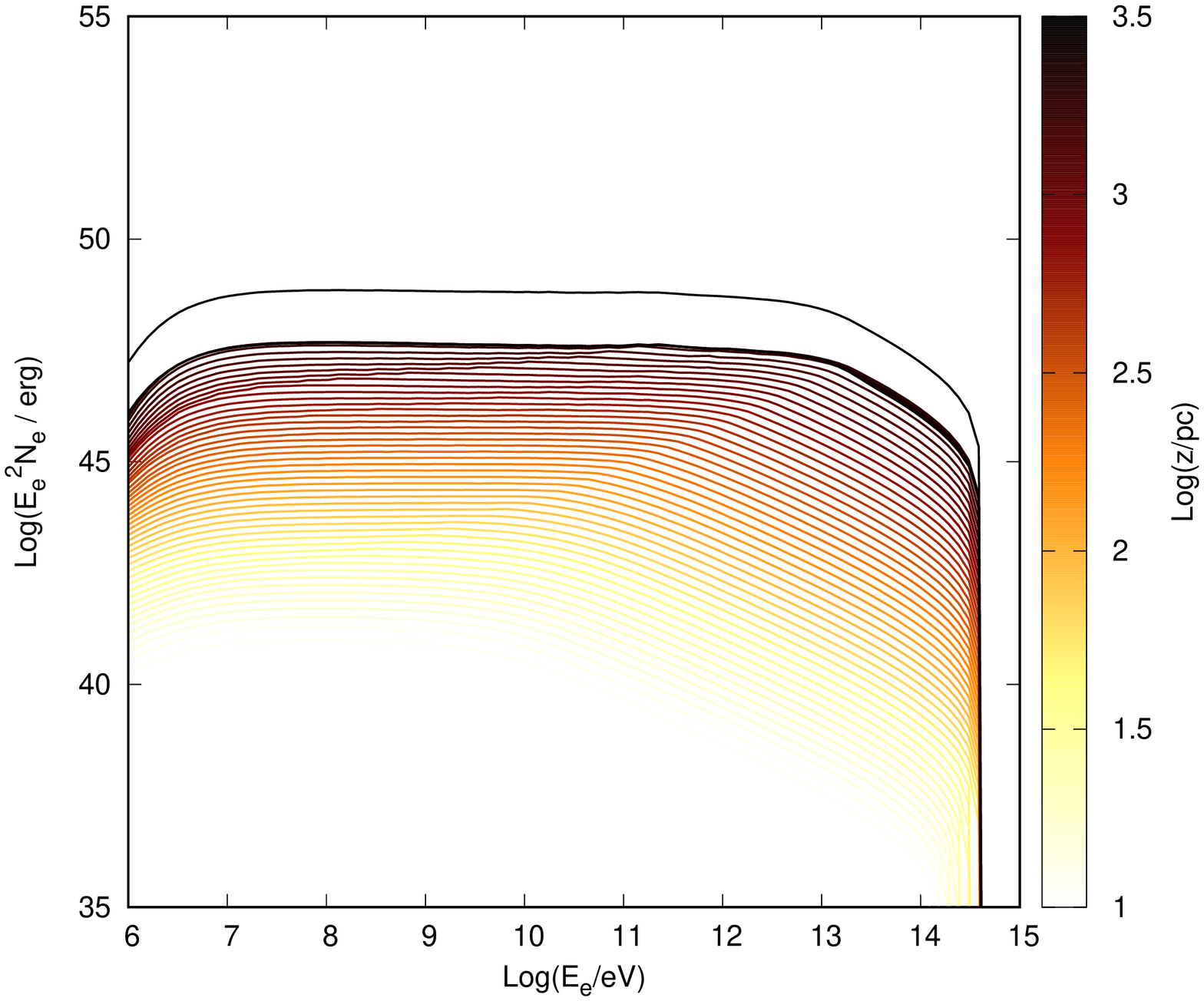} %\hspace{20pt} 
\includegraphics[width=0.3\textwidth,angle=270,keepaspectratio]{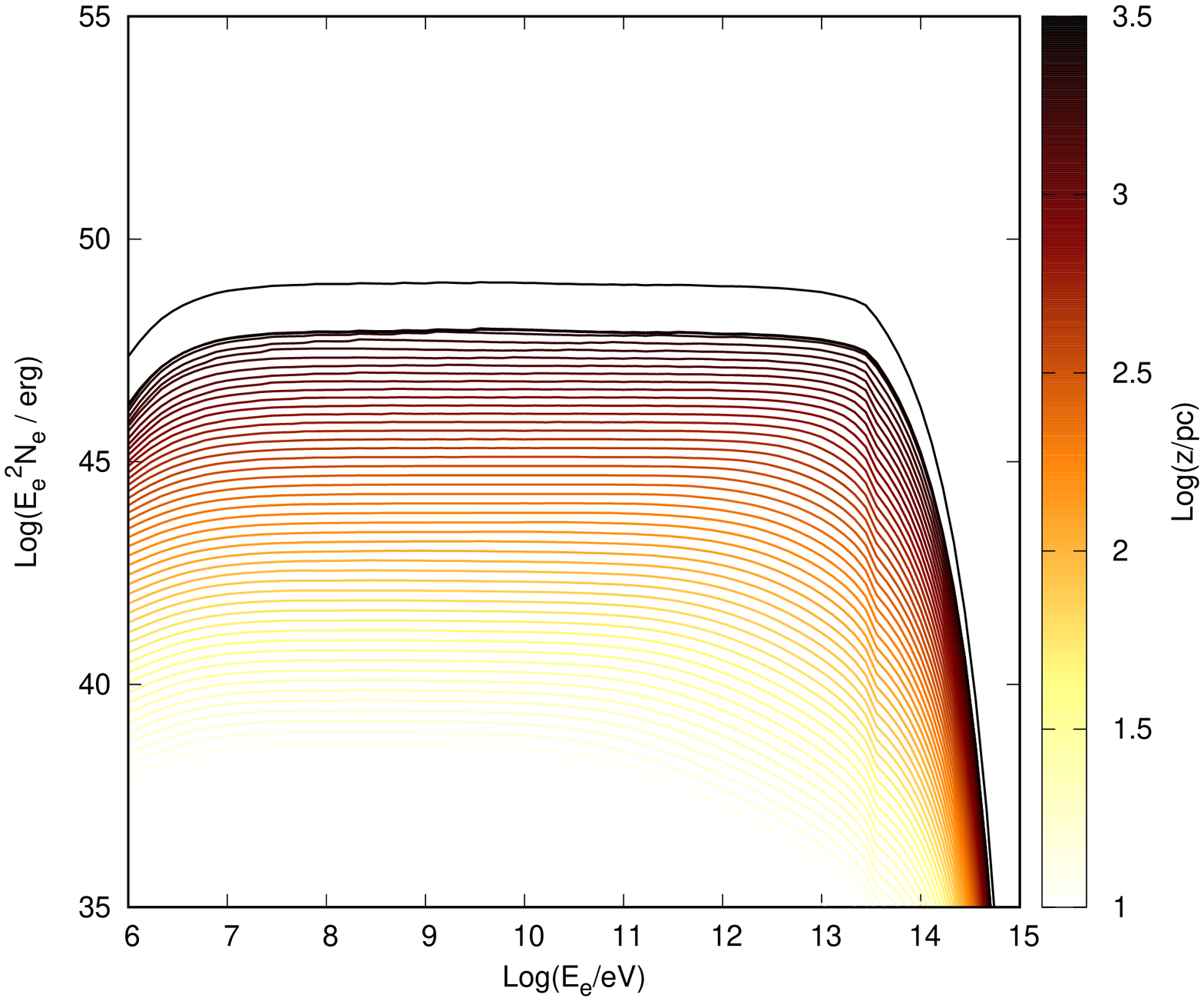}
\vspace{0.5cm}
\caption{Electron energy distribution at different $z$ for M87, in the case $\xi=1$. Left panel is for $B=B_{\rm eq}$, and right panel for $B=10^{-1}B_{\rm eq}$. The black line on top is the integrated distribution.}
\label{fig:M87_dist}
\end{figure*}

Figure~\ref{fig:M87_dist} shows electron energy distributions at different heights obtained for M87 in the case $\xi=1$. For the equipartition magnetic field, the effect of synchrotron cooling is clearly seen in the particle spectra; notice also that particles are able to achieve higher energies at higher $z$. Since the diffusion and acceleration rates vary with $z$ in the same way, the maximum energy is constant along the jet. The non-thermal fraction $\eta_{\rm NT}$ has been fixed to 0.1 as a reference value to compute the emission.

\subsection{Spectral energy distributions}

Once the electron energy distribution is known, we can compute in the flow frame the synchrotron and the IC photon rate per energy unit produced within each height interval ${\rm d}z$: ${\rm d}\dot{N}'_{\gamma}(E'_{\gamma},z)$, and the whole jet SED: 
\begin{equation}
E'_{\gamma}\,L'_{\gamma}(E'_{\gamma}) = \int^{z_{\rm max}}_{z_{\rm min}} \int^{E'_{\gamma\rm max}}_{E'_{\gamma\rm min}} {\rm d}E'_\gamma E'_\gamma {\rm d}\dot{N}_{\gamma}(E'_{\gamma},z)\,,
\end{equation}
where $E'_\gamma$ is the gamma-ray photon energy in the flow frame.

As mentioned above, particles propagate far from the jet-star interaction region, and reach jet scale regions before cooling down significantly. There, particles are advected with the jet velocity, close to $c$, which implies that Doppler boosting must be taken into account \citep{bosch-ramon2015}. In the observer frame, the SED is enhanced according to \citep{lindt1985}:
\begin{equation}%\label{eq:doppler}
E_{\gamma}\,L_{\gamma}(E_{\gamma}) = \delta_{\rm{j}}^4 E'_{\gamma}\,L'_{\gamma}(E'_{\gamma}),
\end{equation}
where $E_{\gamma}=\delta_{\rm{j}} E'_{\gamma}$, and $\delta_{\rm{j}}$ is given by Eq.~(\ref{eq:doppler}).

\section{Results}\label{results}

\begin{figure*}[!ht]
\centering
\includegraphics[width=0.6\textwidth,keepaspectratio]{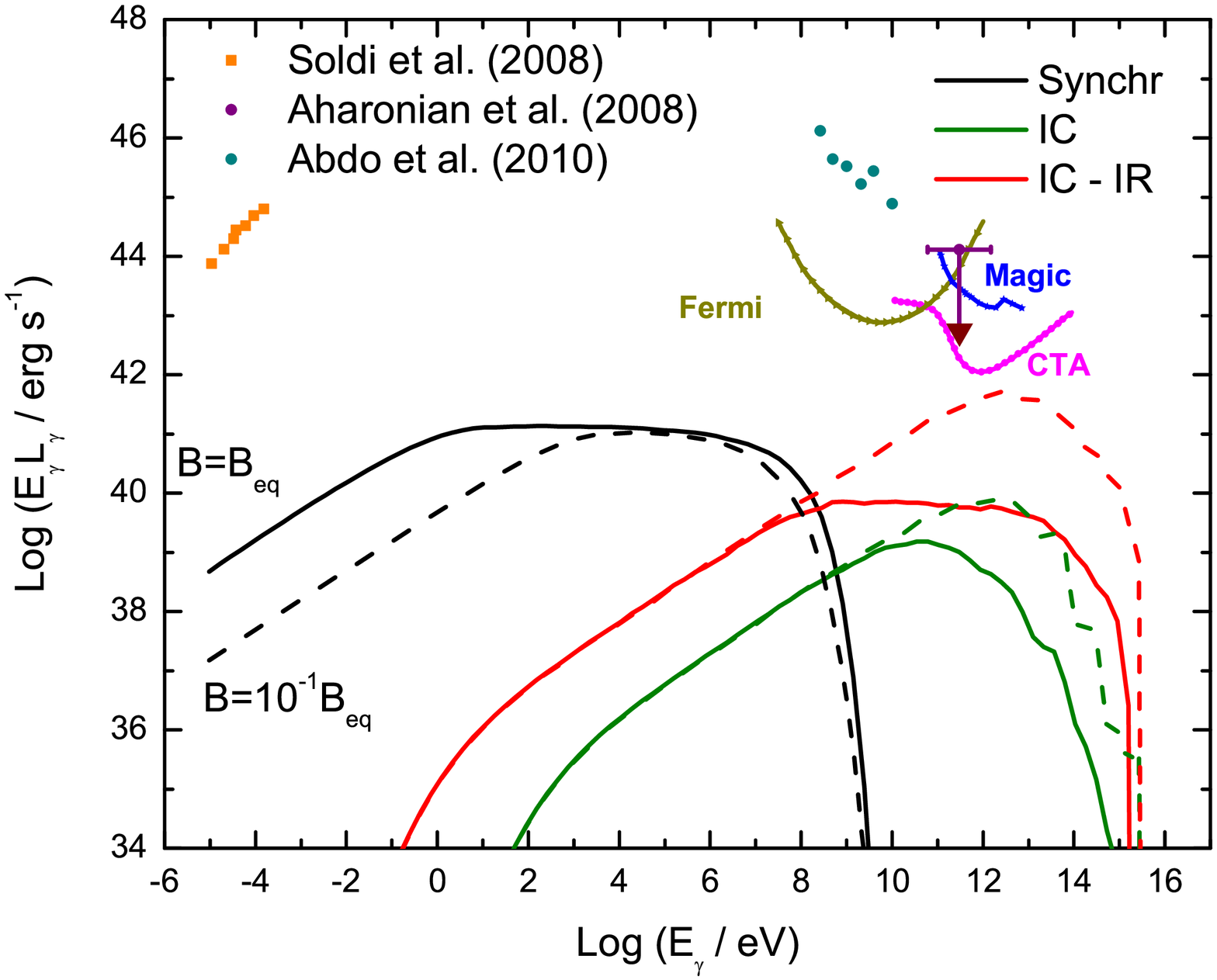} \hspace{20pt} 
\includegraphics[width=0.6\textwidth,keepaspectratio]{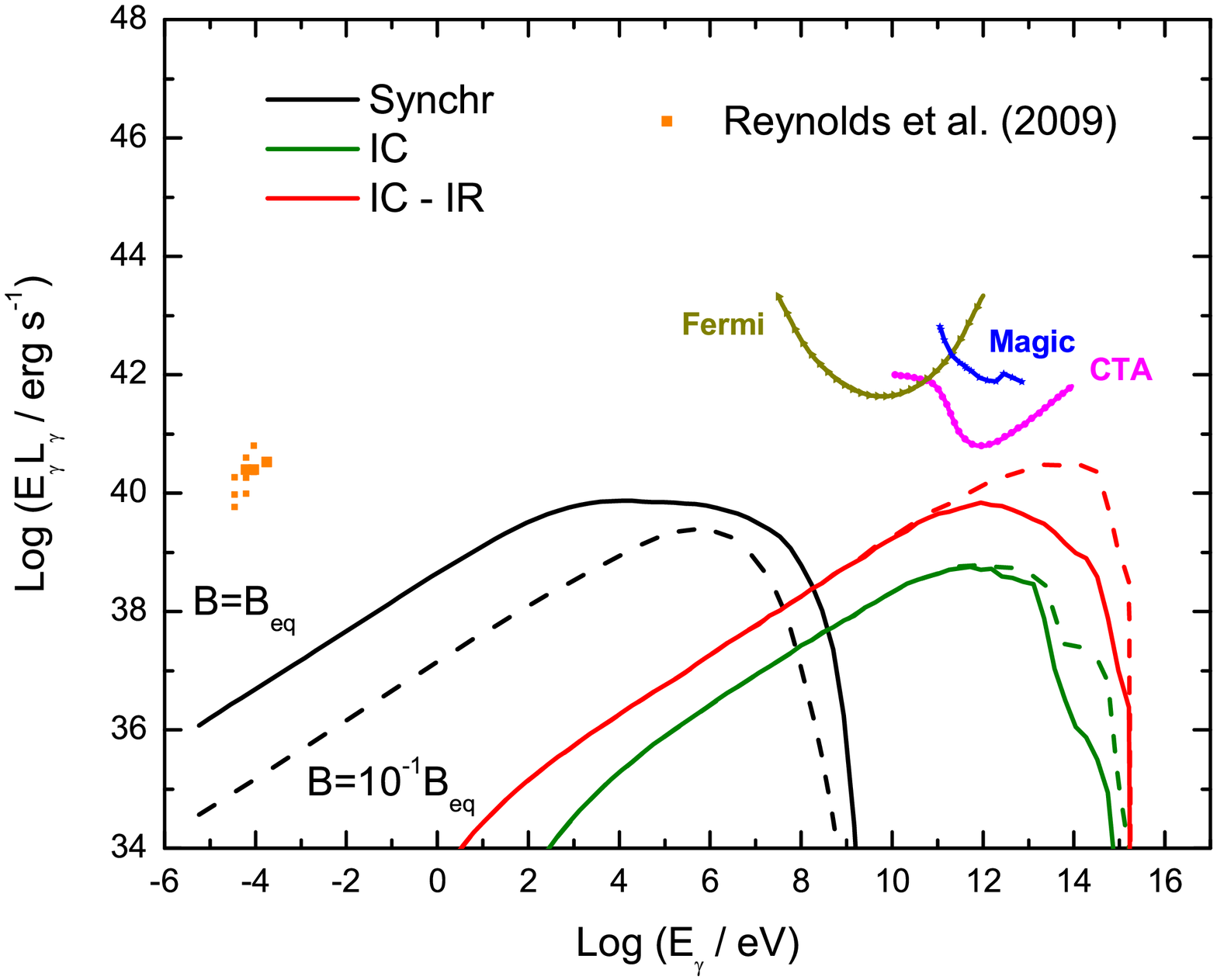}%\hspace{20pt} 
\caption{SEDs of the synchrotron and IC emission for the starburst galaxies. The top panel corresponds to 3C~273, and the bottom panel to Mrk~231. In both panels, solid lines correspond to a magnetic field in equipartition with the jet ram pressure, and dashed lines to the case below equipartition. Radio data for both sources are also presented, along with {\it Fermi} detection and HESS upper limits for 3C~273.}
\label{fig:starburst}
\end{figure*}

Figure \ref{fig:starburst} shows the contribution to the non-thermal luminosity by jet-star interactions on jet scales for the starburst galaxies 3C~273 (top panel) and Mrk~231 (bottom panel). For reference, the figure panels show the sensitivity of three gamma-ray instruments: MAGIC (operating; above $100$~GeV), CTA (forthcoming; above $\sim 30$~GeV), and {\it Fermi} (operating; $\sim 0.1-100$~GeV).

Radio data of 3C~273 and Mrk~231 are also presented in Fig.~\ref{fig:starburst}  \citep{Steenbrugge2010,soldi2008}. In the case of Mrk~231, the radio data were taken between 1996 and 2006, a year in which an intense radio flare was detected \citep[this highly variable radio emission is associated with AGN activity;][]{reynolds2009}. The predicted radio fluxes are well below the typical observed fluxes from Mrk~231 and 3C~273.

For 3C~273, we also show gamma-ray emission detected by {\it Fermi} during a quiescent state in 2009 \citep{abdo2010}, and upper limits to TeV emission by HESS \citep{aharonian2008}. The source is also prominent at X-ray frequencies, although a high percentage in this band comes from a hot corona/disk. Nevertheless, the emission from the interactions studied here does not contribute significantly in X-rays.

\begin{figure*}[!ht]
\centering
\includegraphics[width=0.6\textwidth,keepaspectratio]{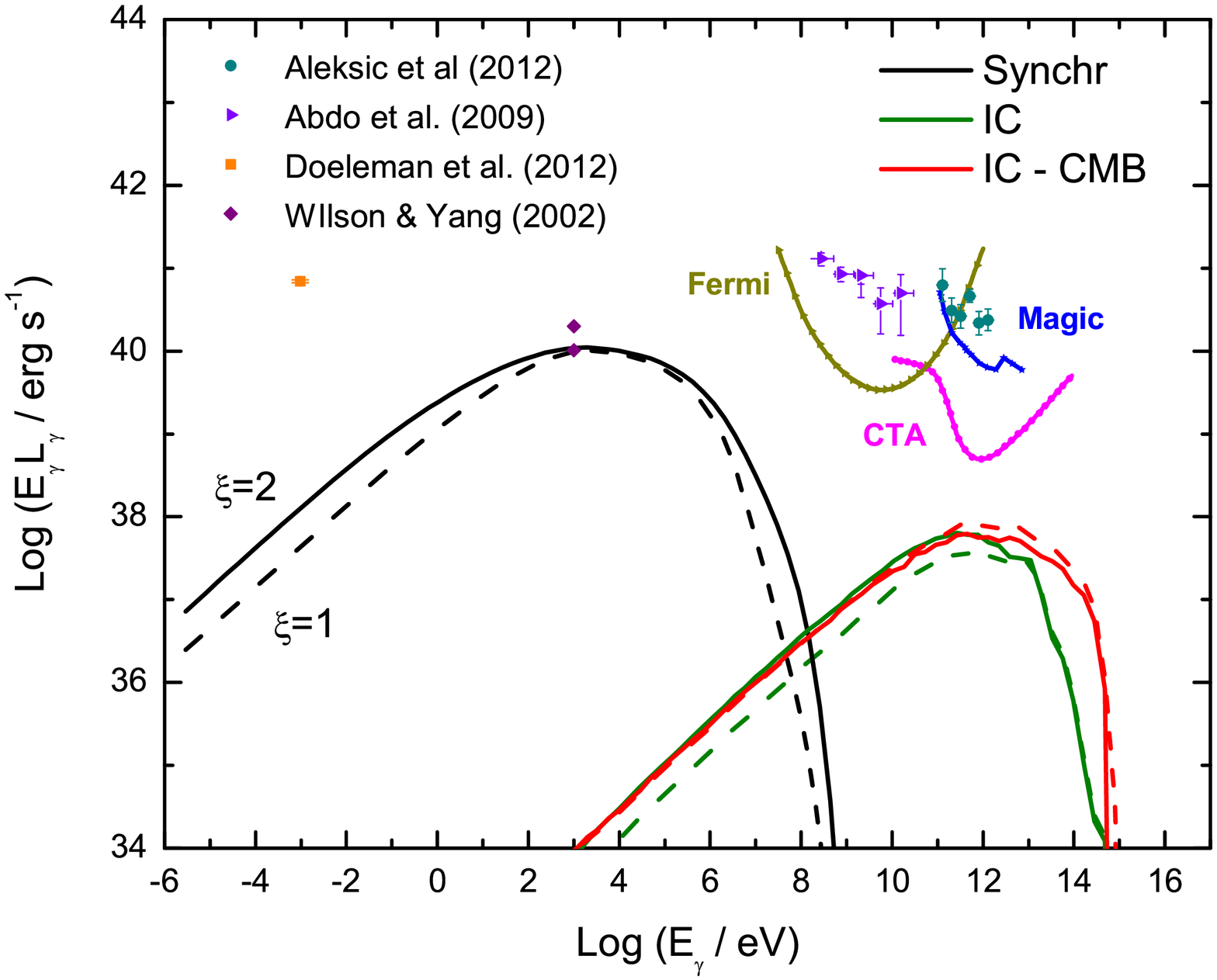}
\includegraphics[width=0.6\textwidth,keepaspectratio]{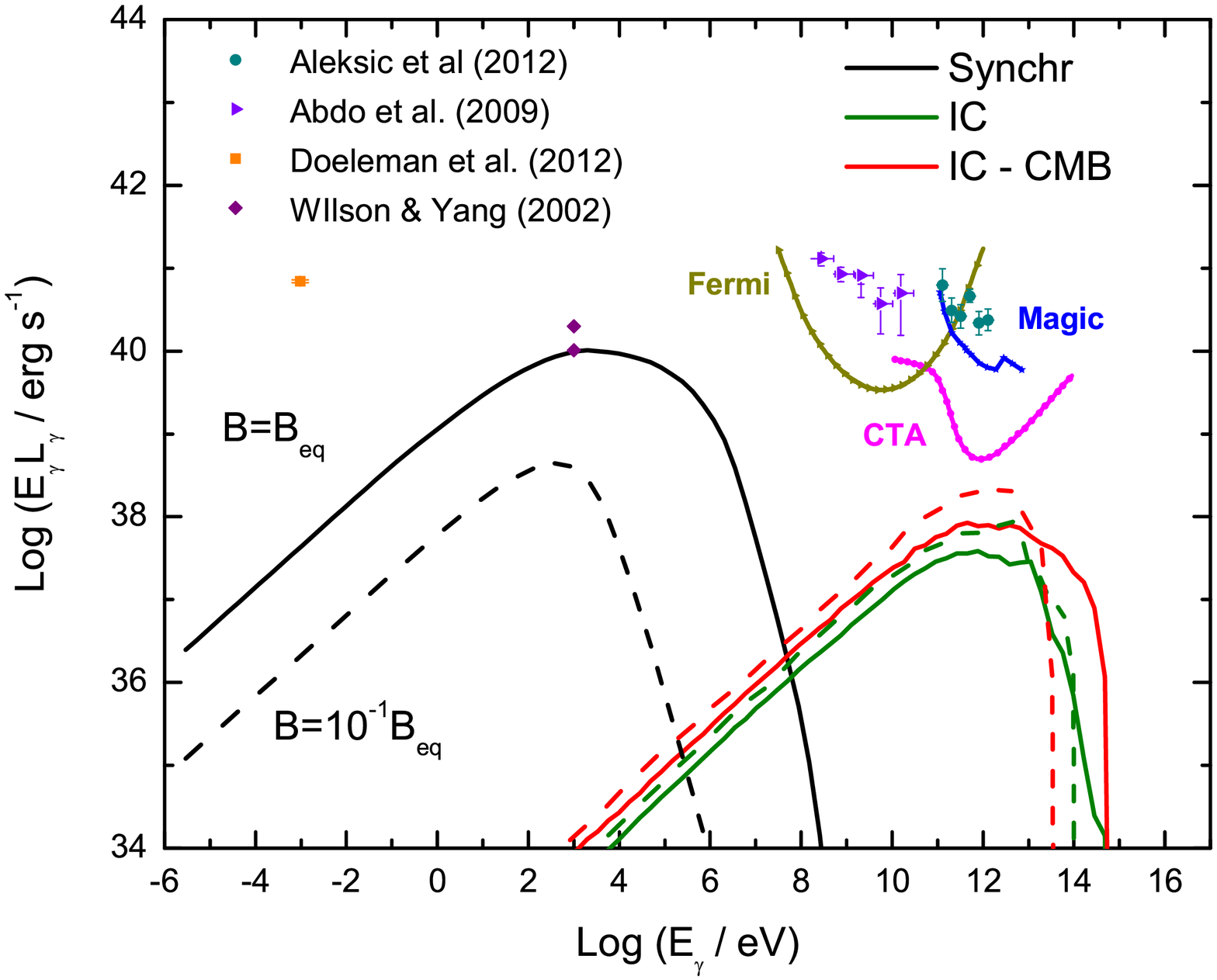}
\caption{SEDs for the jet of M87. The top panel shows the SEDs obtained using  the two stellar indexes. The bottom panel corresponds to a stellar index of $\xi = 1$, and both equipartition and below-equipartition magnetic fields. Sensitivities of gamma-ray detectors are also included, together with the detection by MAGIC \citep{magic2012M87} and {\it Fermi} \citep{abdo2009M87} of M87 during the source steady state. }
\label{fig:M87}
\end{figure*}

In Fig.~\ref{fig:M87} we show the contribution to the non-thermal luminosity of M87 by jet-star interactions on jet scales for the different slopes of the stellar density (top panel), and the two fractions of the equipartition parameter (bottom panel). The figure also shows the sensitivity of the gamma-ray instruments listed above. The data taken by {\it Fermi} correspond to a quiescent state of M87, since during a period of ten months there was no evidence of a flare \citep{abdo2009M87}. Between 2005 and 2007, the MAGIC collaboration collected more than 100~h of observations of M87 in a persistent low-emission state \citep[i.e., no flaring events;][]{magic2012M87}. Both data sets can be used simultaneously, as suggested in \citet{magic2012M87}. The radio luminosity for M87 is from \citet{DoeFis2012}. As in the case of Mrk~231 and 3C~273, there is no conflict between the predicted radio emission and the typical observed fluxes from these sources. 

The inner jet of M87 ($\lesssim 3$ kpc) displays a structure made up of several knots, which can be resolved at radio and optical wavelengths. There is also X-ray emission associated to these knots, but it is slightly shifted upstream with respect to the optical peak \citep{marshall2002}. In addition, the X-ray spectra from the core and the brightest knots (those close to the nucleus) are similar, and the core flux is larger than those predicted by accretion flow models \citep{wilson2002}. All this seems to indicate that an inner jet might contribute (if not dominate) to the X-ray emission. The X-ray fluxes measured close to the nucleus of M87, and in the knot A of the jet, imply luminosities of $\sim 10^{40}$~erg~s$^{-1}$ at 1~keV \citep{wilson2002,marshall2002}. These are similar to the predicted synchrotron luminosity for a magnetic field in equipartition with the jet kinetic power. However, the emission obtained from our model is expected to be diffuse, and unable to reproduce the structure seen in the jet of M87. This suggests two possibilities in our scenario: that a magnetic field below equipartition is a more realistic assumption, or that the acceleration efficiency is lower than the one adopted here.

At TeV energies, the contribution to the luminosity from IC against the CMB is comparable with that against starlight, in agreement with previous results \citep{hardcastle2011}.

\begin{figure*}[!ht]
\centering
\includegraphics[width=0.6\textwidth,keepaspectratio]{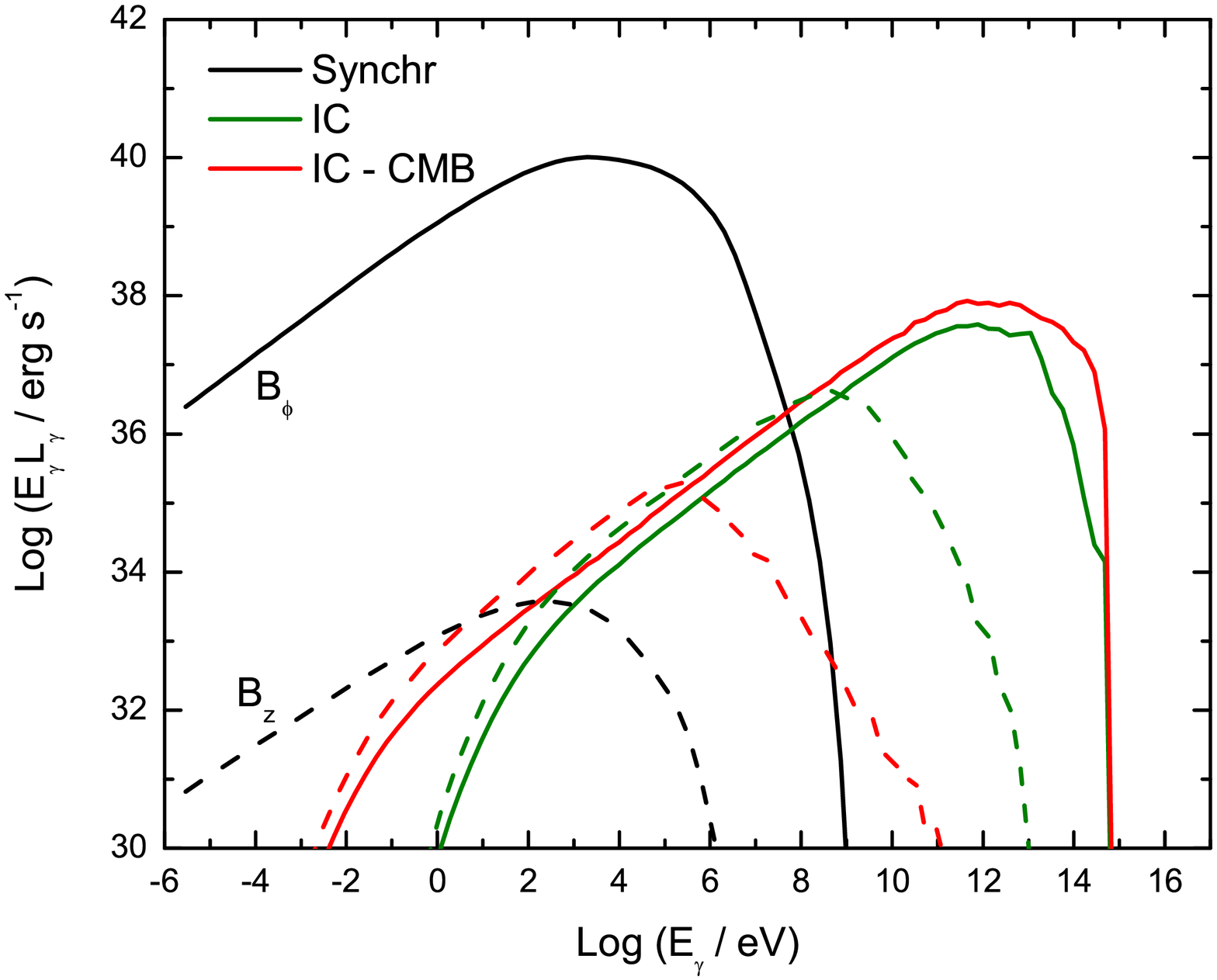}
\caption{SEDs for the jet of M87 in the case of $\xi=1$, produced by the different components of the magnetic fields: poloidal (dashed lines) and toroidal (solid lines). }
\label{fig:M87_B}
\end{figure*}

Although significant transverse components of the magnetic field are found along the jet in M87 (specially in the bright knots), the projected magnetic field lies mostly along the jet \citep{owen1989}. We thus consider a case with a dominant poloidal component. Figure~\ref{fig:M87_B} shows the SEDs obtained for M87 using different configurations of the magnetic field (equipartition values). 

It is worth noting that our jet models miss jet regions larger than those explored where electrons may still radiate through IC in the CMB. This could be particularly relevant for M87 and its disrupted jet regions beyond a few kpc, as its kpc-scale jet emission is already little enhanced by Doppler boosting. A similar effect occurs for 3C~273, where a kpc-scale jet seems to have a larger inclination with respect to the line of sight, hence reducing the Doppler enhancement.

\section{Discussion and summary}\label{discussion} 

In this work, we compute the SEDs of the non-thermal radiation produced by the interaction of extragalactic jets with stars on jet scales. We study two types of galaxy hosts: starburst versus massive elliptical AGN, exemplified by three objects: Mrk~231 and 3C~273 versus M87; for each one, the stellar populations have been characterized. 

For a star-forming galaxy with a high SFR such as Mrk~231, the luminosity in gamma rays computed numerically can be as high as $\sim 10\,\eta_{\rm NT}$\% of the jet luminosity ( $\sim 1\,\eta_{\rm NT}$\% for 3C~273) as long as the magnetic field is a fraction $\lesssim 10^{-2}$ the equipartition value. In that case, the radiation is mostly produced at TeV energies. In all the studied  cases, the radiation comes mainly from the largest scales of the emitter, meaning 100~pc scales for the starbursts, and kpc scales for M87. In M87, as Doppler boosting effects are minor, the non-thermal luminosity reaches only $\sim 0.1\,\eta_{\rm NT}$\% of the jet luminosity. This illustrates the great importance of jet speed and orientation. In addition, the section covered by the stellar winds is smaller than in the case of Mrk~231; not so with respect to 3C~273, but the latter has a much more powerful jet (in addition to the strongest Doppler boosting of the three studied cases). For equipartition fields, unlike Mrk~231 and 3C~273, the synchrotron radiation efficiency in M87 may be significantly higher than for the IC emission, with synchrotron photons reaching $0.1-1$~GeV energies, but most of the emission being released in X-rays. We remark that such high fluxes, expected to be smoothly spatially distributed in our model, are in contradiction with the structured X-ray luminosity observed in the nucleus and knots in M87. Another important difference between M87 and the starburst galaxies is the more diluted target photon fields in the former, as seen when comparing the IC cooling rates in Figs.~\ref{fig:losses_starburst} and \ref{fig:losses_M87}. The available non-thermal-to-jet luminosity ratios obtained in this work range $L_{\rm NT}/L_{\rm j}\sim 10^{-3}-10^{-1}$. These ratios are rather significant, although it is worth pointing out that a defect of stars in the jet directions would proportionally affect $L_{\rm NT}/L_{\rm j}$.

Despite M87 being potentially detectable by CTA for $\eta_{\rm NT}\rightarrow 1$, it seems unlikely that the interactions of its jets with stars on jet scales will contribute significantly to the persistent gamma-ray emission already detected from this source. It does not seem feasible either, given the limitations in angular resolution, to disentangle a putatively detectable, jet-star interaction kpc-scale radiation from other emitting regions of the galaxy center. For the two starburst AGNs studied, in particular for 3C~273, a detection is possible if the magnetic field is well below equipartition and  acceleration efficient. Even Mrk~231 might be detectable with CTA for $\eta_{\rm NT}\rightarrow 1$. Nevertheless, the detectability of these sources ultimately depends on unknown parameters, namely $\eta_{\rm NT}$, and $\alpha$, the latter determining whether gamma rays will be an important radiation channel. Slightly more optimistic Doppler boosting parameters would also significantly improve the detectability of these sources. In summary, the non-thermal emission from jet-star interactions on large scales may represent a non-negligible (persistent) contribution to gamma rays, although the uncertainties are high, and more accurate studies, source-specific or population-based, are still needed to better determine the role of the process at high energies, and constrain the values of the free parameters.

It is worth comparing the global large-scale emission, and the emission emitted locally (close to the interaction region), which has not been calculated in this work \citep[see][]{moreno2016}. To this end, one can compare the radiative efficiency (Eq.~\ref{eq:frad}) at the jet scale to that at the jet-star interaction scale. For a region where escape losses dominate radiative losses, $f_{\rm rad} \sim t_{\rm rad}^{-1} / t_{\rm esc}^{-1} \propto l_{\rm c} w_{\rm ph}$, where $l_{\rm c}$ and $ w_{\rm ph}$ are a characteristic emitter length and the characteristic target photon energy density, respectively. At large scales, $l_{\rm c}\sim z$, and a prescription for $w_{\rm ph}$ is given in Sects.~\ref{appLnt} and \ref{enlo}. Locally, we can approximate $l_{\rm c}\sim 10\,R_{\rm s}(z)$, and $w_{\rm ph}\sim L_{\rm s}/4\pi c (3R_{\rm s}(z))^2$. The interactions of the jet with the most evolved red giants, and with massive stars with $m\gtrsim 40\,M_\odot$, dominate the non-thermal activity; we consider $\dot{M} =10^{-7}\,M_{\odot}$~yr$^{-1}$ and $L_{\rm s}=100\,L_{\odot}$ in the case of red giants, and $\dot{M}=2\times 10^{-6}\,M_{\odot}$~yr$^{-1}$, $v_{\rm w}=10^8$~cm~s$^{-1}$ and $L_{\rm s}=5 \times 10^4\, L_{\odot}$ for massive stars. The temperature of the target photon field also affects the cooling distance of electrons, and has to be included in the analysis; roughly: $f_{\rm rad}^{\rm glob}/f_{\rm rad}^{\rm local}\propto T_{\rm local}/T_{\rm glob}$. Differences in Doppler boosting between the global and the local scales are neglected.

For 3C~273, one obtains:
\begin{equation}
\frac{f_{\rm rad}^{\rm glob}}{f_{\rm rad}^{\rm local}} \sim 1\, \Big(\frac{z}{{\rm pc}} \Big),
\end{equation} 
hence at hundred-pc scale the global IR IC component largely dominates, whereas the global stellar IC component is comparable with the local one. Something similar happens for Mrk~231. For M87, on the other hand, one obtains:
\begin{equation}
\begin{aligned}
\frac{f_{\rm rad}^{\rm glob}}{f_{\rm rad}^{\rm local}} &\approx 2 \times 10^{-3} \Big(\frac{z}{{\rm pc}} \Big),  & \textrm{for } \xi=1, \\
\frac{f_{\rm rad}^{\rm glob}}{f_{\rm rad}^{\rm local}} &\approx 3, \forall z  & \textrm{for } \xi=2\,;
\end{aligned}
\end{equation}  
now, the radiative roles of global CMB and stellar components are comparable, and the temperatures of the dominant target fields are also similar for both the global (taking only red giants) and the local components. Therefore, the global contribution on kpc scales should dominate small-scale contributions for both index values. We recall that the comparison is very crude, and the uncertainty is probably order-of-magnitude.  Nevertheless, the result indicates that the few-kpc scale jet emission from jet-star interactions may easily overcome that from the interaction regions themselves.

The analytic prescription to estimate the apparent luminosity given by Eq.~(\ref{eq:appLnt}),
considering only IC interactions with stellar photons, yields values approximately ten times higher than those derived numerically (almost a hundred times higher for M87). This is somewhat expected, given the crude approximation to compute the radiation efficiency: the actual IC cooling rate at the Thomson-KN transition is slightly below the adopted simple value, and the electrons with lower and higher energies radiate with lower efficiencies. In the Thomson approximation, and for $\alpha=2$, the energy dependence of efficiency already yields an overestimate of the analytical prediction by a factor of $\ln(E'_{\rm IC}/E'_{\rm min})\sim 10$. Therefore, we remark that using the analytical prescription to estimate the gamma-ray luminosity from the jet-star interactions is overestimating its value by approximately a factor of 10.

\section*{Acknowledgments}

We thank the anonymous referee for helpful comments and suggestions. We acknowledge support from the Spanish Ministerio de Econom\'{i}a y Competitividad (MINECO/FEDER, UE) under grants AYA2013-47447-C3-1-P, AYA2016-76012-C3-1-P, with partial support from the European Regional Development Fund (ERDF/FEDER), MDM-2014-0369 of ICCUB (Unidad de Excelencia `Mar\'{i}a de Maeztu'), and the Catalan DEC grant 2014 SGR 86. F.L.V is also supported by the Argentine Agency CONICET (PIP 2014-00338). N.T-A. acknowledges support from MINECO through FPU14/04887 grant. V.B-R. also acknowledges financial support from MINECO and European Social Funds through a Ram\'on y Cajal fellowship. This research has been supported by the Marie Curie Career Integration Grant 321520.

\bibliographystyle{aa}
\bibliography{myrefs7}   %expects file references.bib

\end{document}